%% file: 00_Manuscript.tex
\documentclass[11pt]{article}

\input{./packages/packages.tex}

\input{./packages/Commands.tex}

\usepackage[numbers]{natbib}

\usepackage[a4paper,margin=0.9in]{geometry}
\usepackage{authblk}
\usepackage{hyperref}
\begin{document}

\title{Twisting Kelvin Cells for Enhanced Vibration Control} 

\hypersetup{pdftitle  = {Twisting Kelvin Cells for Enhanced Vibration Control},
	pdfauthor = {Lukas Kleine-Waechter},
	pdfkeywords = {Kelvin cell, Dispersion, Elastic waveguides, band gap engineering, Transmission measurement, SLA printing}
	}

\author[1,3]{Lukas Kleine-W\"achter}
\author[2]{Anastasiia O. Krushynska}
\author[3]{Romain Rumpler}
\author[3]{Huina Mao}
\author[1]{Gerhard M\"uller}
\affil[1]{Chair of Structural Mechanics, Technical University of Munich}
\affil[2]{Faculty of Science and Engineering, University of Groningen} 
\affil[3]{The Marcus Wallenberg Laboratory for Sound and Vibration Research,
KTH Royal Institute of Technology}

\date{}
\maketitle

\begin{abstract}
This study introduces a minimalist design strategy to engineer wave propagation in cellular metamaterials by applying a single-parameter twist to the classical Kelvin unit cell. By breaking the cell's  symmetry while preserving its lattice topology, this geometric modification activates and tunes band-gap frequencies with only a 3\% mass increase, thereby obviating increased geometric complexity and significant mass augmentation often required to achieve 
wave-filtering performance at low frequencies.
The proposed strategy relies on complex-valued Bloch–Floquet analysis of quasi-one-dimensional periodic chains (but is not restricted to) that reveals symmetry-breaking triggers activating two distinct wave attenuation mechanisms: wide Bragg-type and narrow polarization-dependent band gaps arising from longitudinal–torsional mode coupling. These findings are substantiated by a physics-driven analytical model that captures the mode-coupling-induced avoided crossings and provides key insight into the wave characteristics of twisted architectures.
The predictive fidelity of this approach is validated by transmission measurements on SLA three-cell specimens, which reveal up to 20 dB of wave attenuation. We also demonstrate that accounting for material viscoelasticity is essential for accurate prediction, as idealized linear-elastic models are insufficient to capture band-gap frequencies in polymer-based chains. Overall, we proved that modest, well-defined modifications to a classical parent geometry yield advanced wave-filtering performance at low frequencies. This offers a tractable and methodologically complete design strategy for vibration control in lightweight lattices.
\begin{figure}[H]
\centering
  \includegraphics[width=0.9\columnwidth]{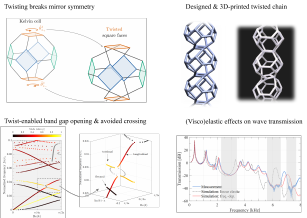}
\end{figure}
\end{abstract}

%
\input{./content/01_Intro}
\input{./content/02_UnitCell}

\input{./content/03_DispersionAnalyses}
\input{./content/04_Experiments}
\input{./content/05_Conclusion}
%
\section*{Acknowledgements}
The manufacturing and measurements were carried out at the Engineering and Technology Institute Groningen of the University of Groningen. The authors gratefully thank Shantanu Nath of the \textsc{Metamechanics} group for providing access to the laboratory facilities and for his continuous support and supervision during the experimental campaign. The authors also thank Stefan Jacob (KTH) for his valuable input in planning the measurement setup. Financial support from the Swedish Research Council (VR Grant No. 2021--05791) is gratefully acknowledged.

\section*{Conflicts of Interest}
The authors declare no known conflicts of interest.

\bibliographystyle{unsrt}
\bibliography{References}

\clearpage
\appendix
\renewcommand{\theequation}{\thesection.\arabic{equation}}
\setcounter{equation}{0}
\renewcommand{\thefigure}{\thesection.\arabic{figure}}
\renewcommand{\thetable}{\thesection.\arabic{table}}

\setcounter{figure}{0}
\setcounter{table}{0}
\setcounter{equation}{0}
\input{./content/06_a_Appendix_Methods}

\clearpage
\setcounter{equation}{0}
\input{./content/06_b_Appendix_AnalyticalModel.tex}
\clearpage
\setcounter{equation}{0}
\input{./content/06_c_Appendix_Experiments.tex}

\end{document}

%% file: packages/packages.tex
\usepackage{mfirstuc}
\usepackage[utf8]{inputenc}
\usepackage{longtable,tabularx}
\usepackage{amsmath,amssymb,siunitx,epstopdf,pdfpages,bm,relsize, mathtools}
\usepackage{pdfpages}
\usepackage{eurosym}
\usepackage{graphicx}
\graphicspath{{Figures/}}
\usepackage{tikz}
\usepackage{svg}
\svgpath{{svg/}}
\usepackage{pgfplots} 
\usepackage{pgfgantt}
\usepackage{pdflscape}
\pgfplotsset{compat=newest} 
\pgfplotsset{plot coordinates/math parser=false}
\newlength\fwidth		

\usepackage{etoolbox}
\patchcmd{\paragraph}{\itshape}{\normalfont\bfseries}{}{}
\usepackage{caption}
\usepackage{subcaption}
\usepackage{capt-of} 
\usepackage{floatrow}
 \usepackage{enumitem}
\usepackage{microtype}
\usepackage{subcaption}
\usepackage{cmap}
\usepackage{booktabs}
\usepackage[english]{varioref}
\usepackage{nicefrac}
\usepackage{scalerel,stackengine}

\usepackage{xcolor}
\definecolor{ml}{RGB}{250,245,235} 
\definecolor{TUM}{RGB}{0,101,189} 

\usepackage[most]{tcolorbox}
\definecolor{mlbrown}{RGB}{181,140,104} 
\definecolor{mlcream}{RGB}{250,245,235}  

\tcbset{
  beerbox/.style={
    colback=mlbrown,
    colframe=mlbrown,
    coltext=black,
    boxrule=0pt,
    arc=4pt,
    left=6pt,
    right=6pt,
    top=6pt,
    bottom=6pt
  }
}

\usepackage[numbers,sort&compress]{natbib}

%% file: packages/commands.tex
\let\otextsc\textsc
\renewcommand{\textsc}[1]{\mbox{\otextsc{#1}}}



%% file: content/01_Intro.tex
Lattice metamaterials occupy a distinct niche within the broader metamaterials landscape~\citep{Phani.2017, Schaedler.2016, Jiang.2022, Ma.2025}, as their ligament-based architectures enable a wide design space governed by ligament type, orientation, and connectivity. This geometric tunability gives rise to unusual macroscopic responses, including vanishing shear stiffness~\citep{Kadic.2012, Buckmann.2014, Brambilla.2025}, negative Poisson’s ratios~\citep{Wang.2020, Zhang.2025b}, and programmable anisotropy~\citep{Mao.2020b, Fu.2025}.

In the dynamic regime, this design flexibility extends naturally to the control of wave propagation. Here, the analysis shifts from macroscopic averages to the unit-cell scale, typically described through the dispersion relation, where the architecture of the unit cell is engineered to shape the band structure and resulting wave propagation characteristics. The symmetry and topology of the unit cell govern the eigenfrequency distribution and their organization in wavenumber space, giving rise to frequency band gaps -- spectral regions in which elastic wave propagation is inhibited~\citep{Brillouin.1953, Hussein.2006, Craster.2023}. These band gaps are traditionally attributed to either Bragg-type destructive interference or local resonances suppressing propagating waves by stationary modes of internal substructural elements~\citep{Liu.2000, Krodel.2014, Junyi.2016, Krushynska.2017, Aguzzi.2022}. 

\begin{figure}[]
\centering
    \begin{subfigure}[b]{0.3\textwidth}
    \centering
    \includegraphics[width=.65\textwidth]{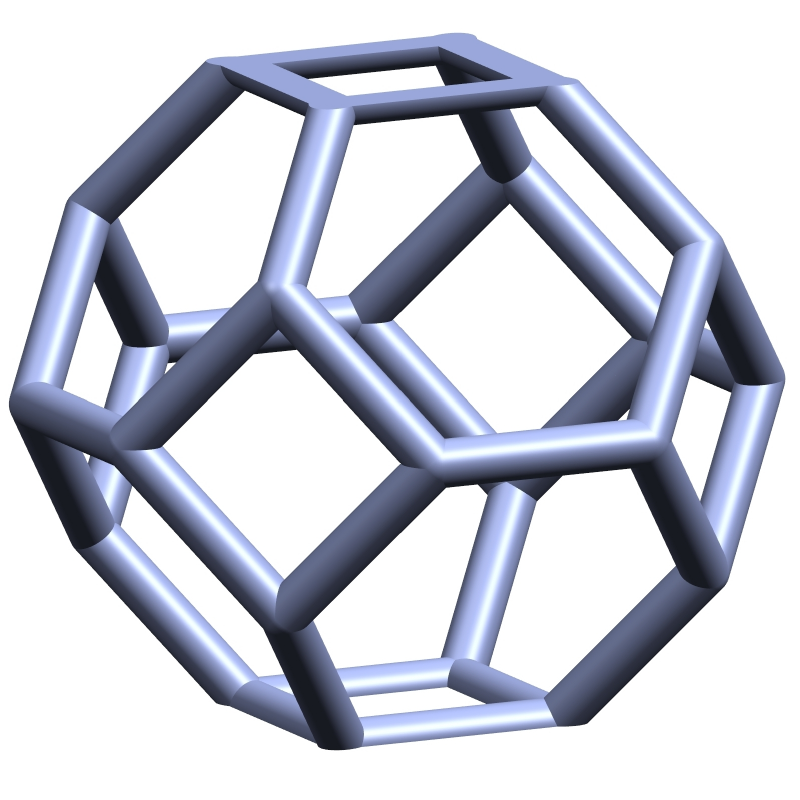}
    \caption{}\label{fig:2_CAD_KC}
    \end{subfigure}
    \hfill
    \begin{subfigure}[b]{0.3\textwidth}
\centering
\includegraphics[width=.65\textwidth]{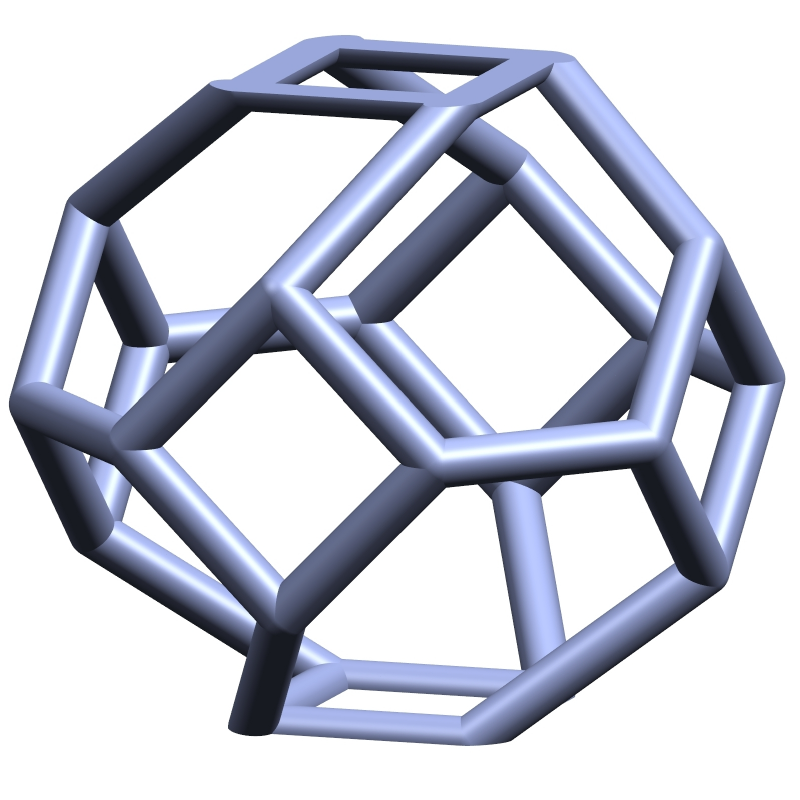}
\caption{}\label{fig:2_CAD_Twist45}
    \end{subfigure}
    \hfill
    \begin{subfigure}[b]{0.3\textwidth}
    \centering
    \includegraphics[width=.65\textwidth]{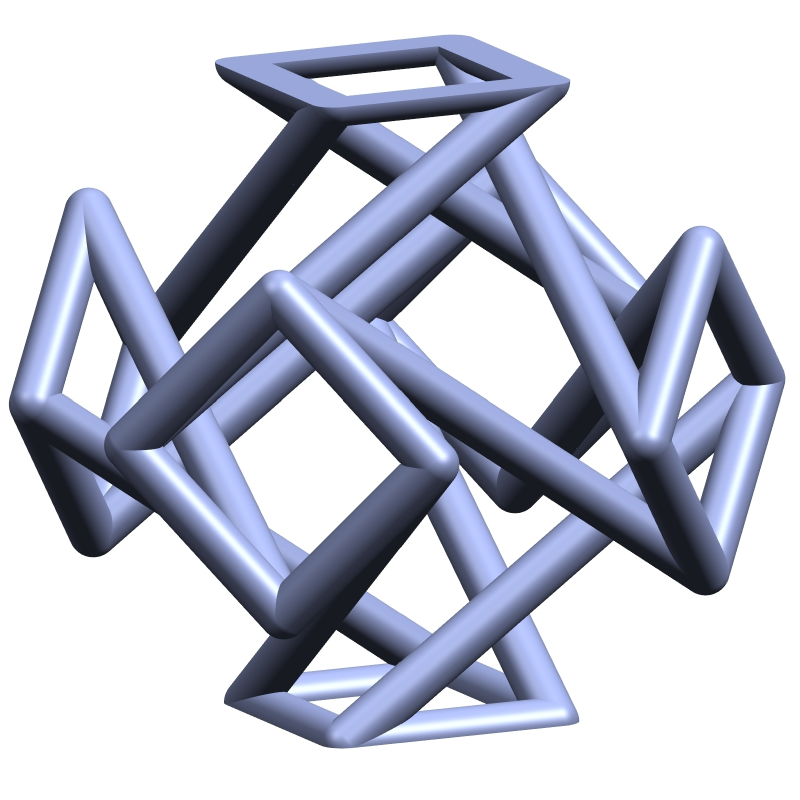}
    \caption{}\label{fig:2_CAD_fullTwist90}
    \end{subfigure}
\caption{Conceptual approach to symmetry breaking in cellular periodic lattices:
 (a)~The reference achiral Kelvin unit cell; 
 (b)~unit cell with a single axial twist; and 
 (c)~a triple-axial twist configuration. 
 This topology-preserving transformation breaks the initial octahedral symmetry of the Kelvin cell to trigger coupled wave mechanics without increasing structural complexity. Note that the halves of the top and bottom square faces are omitted to facilitate tessellation into periodic chains (periodicity constant $a=h_c$, as detailed in Fig.~\ref{fig:2_Sketch_TwistingApproach}).}\label{fig:2_CAD_TwistingApproach}
\end{figure}

Despite this capability of architected lattices to generate large band gaps, a fundamental challenge persists in balancing dynamic performance with structural efficiency: maximizing band gap width often incurs significant mass penalties or reduction in static stiffness~\citep{Liu.2018, Liu.2021}. Rather than addressing this trade-off through increasingly intricate topologies, we instead identify the minimal degree of asymmetry required to induce band gap behavior. To this end, we adopt the canonical Kelvin cell -- an idealized representation of open-cell foams with high initial octahedral symmetry~\citep{Thomson.1887, Gibson.2003, Spadoni.2014} -- as a baseline to isolate symmetry-breaking mechanisms within a single lattice unit. We show that a topology-preserving twist of the square faces (Fig.~\ref{fig:2_CAD_TwistingApproach}) activates coupled axial-torsional interactions. When tessellated into syndiotactic supercells, these interactions give rise to both Bragg-type and polarization-dependent band gaps, yielding up to 20 dB attenuation in a finite chain of only three unit cells. This minimal geometric modification, which increases mass by just 3\%, enables enhanced wave control while preserving topological simplicity and manufacturing.

By prioritizing simplicity and manufacturability, our approach provides a deliberate counterpoint to a prevailing direction in lattice metamaterials research, where symmetry breaking is often achieved through increasingly complex topologies to realize ultra-wide band gaps. A central theme in this context is the use of chiral architectures, which have been shown to exhibit enriched dynamic behavior~\cite{Frenzel.2017, Wu.2018, FernandezCorbaton.2019, Chen.2025, Lee.2024, Montazeri.2025b}. Such structures enable coupled deformation states, including axial-transverse or axial-torsional interactions~\cite{Li.2023, Montazeri.2025, Yves.2026}, that remain decoupled in highly mirror-symmetric configurations, and can be harnessed to generate inertial amplification band gaps~\cite{Yilmaz.2007, Frandsen.2016, Orta.2019}. Within this framework, tacticity, defined here as the controlled modulation of structural handedness, has emerged as an effective strategy to induce band gaps via motion cancellation across unit-cell boundaries, often implemented through syndiotactic supercells~\citep{Bergamini.2019, Ding.2022, Zhao.2022b, Ding.2024b}. 

While effective in maximizing spectral bandwidth, the prevailing trend toward increasingly intricate topologies introduces substantial functional and manufacturing constraints that our work seeks to mitigate. This direction often combines inertial amplification with local resonance in chiral-symmetric architectures~\cite{Zhang.2024, Ding.2024, Yu.2025, Zhang.2025d, Xu.2026}, or exploits structural hierarchy, such as embedding chiral micro-resonators within a host frame, to broaden attenuation regimes~\cite{Yuan.2025}. Additional tunability is frequently achieved through non-uniform ligament geometries~\cite {Chen.2017, He.2025, Gao.2025, Li.2025b} or nonlinear stiffness mechanisms~\cite{Yan.2022, Bai.2025}, often requiring multi-material implementations~\cite{Zhang.2026}. These approaches are commonly paired with topology optimization or machine learning frameworks that prioritize theoretical wave attenuation~\cite{Jiang.2022b, Lee.2024b, Tikani.2025, Cool.2025, Zhang.2026b}, while practical considerations such as manufacturability and robustness remain comparatively unexplored~\cite{Zhang.2024b, Park.2025, Turlin.2026}.

In contrast, our twisted Kelvin cell architecture directly addresses this trade-off by identifying the minimal degree of structural asymmetry required to achieve enhanced wave attenuation performance, favoring a geometrically simple, topology-preserving symmetry-breaking strategy over increasing structural complexity. Building on the established potential of Kelvin cell-based lattices for static anisotropy~\citep{Spadoni.2014, Mao.2020, Mao.2020b, Wei.2023} and vibroacoustic tunability~\cite{Bayat.2018, Bayat.2019, Rice.2020, Cai.2025, Chen.2021}, we leverage this transformation to determine the critical twist necessary to activate band gaps through structural chirality. In contrast to hierarchical or curved architectures that complicate fabrication and increase sensitivity to manufacturing tolerances~\cite{Fabro.2020b, Zhang.2023, Sachdeva.2025b}, the monolithic nature of the twisted Kelvin cell preserves fabrication simplicity while maintaining a direct link between symmetry modification and wave attenuation. Moreover, whereas designs incorporating bulky masses connected by slender chiral ligaments~\cite{Bergamini.2019, Zheng.2017, Zhao.2022, Park.2022, Ding.2022, Ding.2023, Liu.2024, Zhang.2025b, Yu.2025, Yu.2026} often suffer from reduced static strength and fatigue resistance due to stress concentrations at junctions~\cite{Zaccherini.2021}, our approach preserves structural integrity without introducing dead-weight vulnerabilities. Unlike many existing chiral architectures that are restricted to bespoke, unidirectional chains, the proposed twisting mechanism is inherently multi-axial: although the present study focuses on one-dimensional (1D) chains for clarity, the concept extends naturally to all three principal directions, enabling fully three-dimensional (3D) chiral lattices with multi-directional response.

Methodologically, this work departs from the conventional emphasis on the real-valued dispersion relation, which neglects evanescent wave characteristics and the coupling information encoded in the imaginary component of the wavenumber~\cite{Nadejde.2025, Laude.2020}. Instead, we adopt a complex-valued $k(\omega)$ formulation to capture the longitudinal-torsional mode coupling induced by symmetry breaking, supported by an analytical model that provides physical insight into the underlying mechanics. By resolving the full dispersion spectrum, including both propagating and evanescent modes, we obtain a more complete description of wave attenuation than approaches limited to purely real Bloch solutions.

Finally, we address a significant gap in the experimental validation of additively manufactured lattice metamaterials. While prior studies often confirm the existence of band gaps, they frequently overlook the accurate prediction of band-gap boundaries, which can shift significantly under idealized linear-elastic assumptions~\cite{Krushynska.2016, DAlessandro.2018, Krushynska.2021, Krushynska.2021b}. In Section~\ref{sec:experiments}, we demonstrate that neglecting viscoelastic material behavior leads to systematic discrepancies in these predictions. By incorporating the intrinsic viscoelasticity of polymer-based materials, we show that accurate modeling of wave transmission requires accounting for such effects. Through extensive experimental realization of twisted Kelvin cell chains, we prove that integrating viscoelastic material behavior with minimal symmetry breaking enables robust and predictable wave control.

%% file: content/02_UnitCell.tex
\section{Kelvin cell-based lattice architectures}\label{sec:02_Unit}
\begin{figure}[]
    \begin{subfigure}[b]{0.33\textwidth}
         \centering
         \includegraphics[width=.95\textwidth]{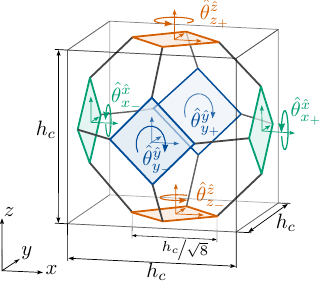}
         \caption{}\label{fig:2_Elementary_KC}
     \end{subfigure}
    \begin{subfigure}[b]{0.33\textwidth}
         \centering
         \includegraphics[width=.95\textwidth]{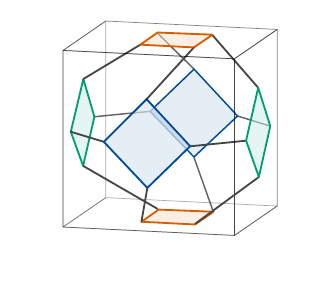}
         \caption{}\label{fig:2_KC_Twist45}
     \end{subfigure}
         \begin{subfigure}[b]{0.33\textwidth}
         \centering
         \includegraphics[width=.95\textwidth]{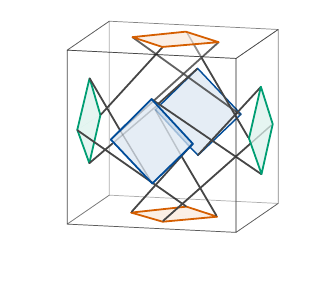}
         \caption{}\label{fig:2_KC_TwistFull90}
     \end{subfigure}
    \caption{Illustration of the twisting operation.
    (a)~The reference (achiral) configuration of the Kelvin cell with the local coordinate system and rotation angles~$\theta_j^{\hat{i}}$ defining the twisting of the square faces.
    (b)~The configuration with one axial twist obtained by rotating the top and bottom square faces about the local $\hat{z}$-axes in opposite directions by~$\theta_{z{+}}^{\hat{z}} = -\,\theta_{z{-}}^{\hat{z}} = 45^\circ$.
    (c)~The configuration with three axial (chiral) twists with all square faces rotated by~$\theta_j^{\hat{i}} = \pm 90^\circ$ in opposite directions.
    (a)-(c) The connectivity pattern is preserved in all cases, while the ligament orientations are modified. Colors are for visualization only.}\label{fig:2_Sketch_TwistingApproach}
\end{figure}

An adjustable unit-cell approach, as originally proposed by some of the co-authors, is adopted to facilitate controllable modifications of the microstructural geometry, treating the \text{Kelvin} cell as a geometric template~\cite{Mao.2020b}. The basic geometric structure, shown in Fig.~\ref{fig:2_Elementary_KC}, consists of three pairs of opposing square surfaces along the three principal directions. Connecting adjacent square faces creates a network of 36 ligaments that form a basic cellular unit. In this configuration, the cell exhibits octahedral symmetry and a uniform cell height~$h_c$ along the three principal directions. 

Geometrical modifications of the described geometry can be achieved through coordinate transformations that describe pairwise in-plane rigid body rotations of the square faces, as illustrated in Fig.~\ref{fig:2_Sketch_TwistingApproach}. Depending on whether the faces are rotated in the same or opposite directions, the resulting configuration remains achiral or becomes chiral. Note that the rotations only alter the positions of the ligament vertices, not affecting the connectivity of the ligaments. Therefore, the modified structures are topologically equivalent to the reference \text{Kelvin} cell, and all twisted faces remain in their original planes, retaining the simple-cubic periodicity in the rotated lattice. 

Mathematically, the \textit{twisting}-operation is described by transformations between the cell's local and global coordinate systems using Euler rotation angles $\theta_j^{\hat{x}}, \theta_j^{\hat{y}}, \theta_j^{\hat{z}}$. The sets of local coordinates $\left( \hat{x},\hat{y},\hat{z} \right)_j$ originate in the center of the respective square faces denoted $j$ and are parallel to the cell's global coordinates, as shown in Fig.~\ref{fig:2_Elementary_KC}. Assuming successive, independent rotations around the axes $\hat{z}, \hat{y}$ and $\hat{x}$, the rotation matrices read \par\nobreak
\vspace{-4mm}
{\small
\begin{flalign}
\begin{aligned}
& \left[R\right]_j =  
   &\begin{bmatrix}
    1 & 0 & 0 \\
    0 & \mathrm{c.}{\theta_j^{\hat{x}}} & \pm \mathrm{s.}{\theta_j^{\hat{x}}} \\
    0 & \mathrm{s.}{\theta_j^{\hat{x}}} & \mathrm{c.}{\theta_j^{\hat{x}}} \\
    \end{bmatrix}
    \begin{bmatrix}
    \mathrm{c.}{\theta_j^{\hat{y}}} & 0 & 
    \mathrm{s.}{\theta_j^{\hat{y}}} \\
    0 & 1 & 0 \\
    \pm \mathrm{s.}{\theta_j^{\hat{y}}} & 0 & \mathrm{c.}{\theta_j^{\hat{y}}} \\
    \end{bmatrix} 
        \begin{bmatrix}
    \mathrm{c.}{\theta_j^{\hat{z}}} & \pm \mathrm{s.}{\theta_j^{\hat{z}}} & 0 \\
    \mathrm{s.}{\theta_j^{\hat{z}}} & \mathrm{c.}{\theta_j^{\hat{z}}} & 0 \\
   0 & 0 & 1 \\
    \end{bmatrix}
    \end{aligned},
    \label{eq:TwistingApproach}
\end{flalign}
}%
with $\mathrm{c}.\cdot = \cos(\cdot)$ and $\mathrm{s}.\cdot = \sin(\cdot)$. Consequently, the location of points $\left(\hat{x},\hat{y},\hat{z} \right)_j^T $ at the square face $j$ of the KC reference lattice is given by
\begin{flalign}
\left(\hat{x}',\hat{y}',\hat{z}' \right)_j^T =  \left[R\right]_j \left(\hat{x},\hat{y},\hat{z} \right)_j^T ,
\end{flalign}
after twisting the reference structure.  The rotation angle and its direction represent additional parameters that alter the unit cell, complementary to the cell height, $h_c$, and the ligament diameter, $d$. While the twist angle affects the inclination of ligaments connecting the square faces, the direction of rotation impacts the unit cell's mirror symmetries. In particular, twisting all square faces in opposing directions creates chiral lattice configurations, as exemplified in Fig.~\ref{fig:2_KC_Twist45} and~Fig.~\ref{fig:2_KC_TwistFull90}. Using the transformed coordinate set and maintaining the connectivity pattern allows the generation of the unit-cell architectures shown in Fig.~\ref{fig:2_CAD_TwistingApproach}, with ligament diameter~$d$ and periodicity constant~$a$. In the 1D-periodic cases discussed in this contribution, the periodicity constant~$a$ is equivalent to $h_c$. Importantly, the twisting operation preserves the characteristic length of the unit cell, while the ligaments attached to the rotated square faces experience an increase in length. Adjusting the alteration of the square faces independently in the manner described provides the opportunity to create lattices with significant and controllable degrees of anisotropy in the resulting elastic properties~\cite{Mao.2020}.

%% file: content/03_DispersionAnalyses.tex
\section{Wave dispersion in infinite designs}\label{sec:Dispersion}
The effects of twisting the unit cell geometry on their wave propagation properties are first studied under the assumption of an infinitely extended medium. For this, the structures are formed by periodic tessellations of Kelvin cells along a single direction, and the calculated dispersion relation captures their \textit{free} wave propagation characteristics.

\subsection{Dispersion relation -- Band structure calculation}
Assuming linear-elastic isotropic homogeneous material behavior, wave propagation in the Kelvin cells lattice is governed by the elastodynamic \text{Navier-Cauchy} equation,
\begin{align}
    \frac{E}{2(1+\nu)(1-2\nu)}\nabla (\nabla \cdot \bm{u}) + \frac{E}{2(1+\nu)} \nabla^2 \bm{u} = -\rho\omega^2 \bm{u},
    \label{eq:Cauchy}
\end{align}
in terms of the displacement vector~$\bm{u} = [u_x, u_y, u_z]^\mathrm{T}$ with a time-harmonic dependency for angular frequency~$\omega$. The other parameters are
the \text{Young’s} modulus~$E$, the mass density~$\rho$, and the \text{Poisson’s} ratio~$\nu$ of a constituent material. 
The lattice periodicity enables reducing the computational domain to a single representative unit cell~$\Omega_\mathrm{U}$ using the \text{Bloch-Floquet} theory~\cite{Floquet.1883}, which results in the following form of the solutions to Eq.~\eqref{eq:Cauchy}:
\begin{align}
    \bm{u}\left(\bm{r}, \bm{k} \right) = \tilde{\bm{u}}_{\bm{k}}\left(\bm{r}, \bm{k} \right) \mathrm{e}^{\mathrm{i}\bm{k}\cdot\bm{r}} \;\mathrm{in}\;\Omega_\mathrm{U},  
    \label{eq:Bloch}
\end{align}
with $\bm{k}$ denoting the wave vector, $\bm{r}$ indicating the position vector in a lattice, and $\tilde{\bm{u}}_{\bm{k}}$ the \text{Bloch} functions that share the lattice periodicity, i.e., $\tilde{\bm{u}}_{\bm{k}} (\bm{r}) = \tilde{\bm{u}_{\bm{k}}} (\bm{r}+ \bm{R})$, where $\bm{R}$ is a \text{Bravais} lattice vector. As we consider the periodicity only along a single dimension, the wave vector takes the form of $\bm{k} = [0, 0, k_z]^T$ and Eq.~(\ref{eq:Bloch}) can be rewritten as
\begin{align}
    \bm{u}(\bm{r}) = \bm{u}(\bm{r} + \bm{a})\,\mathrm{e}^{\mathrm{i} \bm{k}\cdot\bm{a}} \;\;\mathrm{on}\;\mathrm{\Gamma_{\Omega_\mathrm{U}}},
    \label{eq:BC}
\end{align}
with $\bm{a} = [0, 0, a]$, where $a$ is the lattice period. All remaining structural boundaries are treated as traction-free. Substituting Eq.~\eqref{eq:Bloch} and 
Eq.~\eqref{eq:BC} in wave equation~\eqref{eq:Cauchy}, one obtains an eigenvalue problem; the solutions of which provide a dispersion relation:
\begin{align}
    \left[ \mathbf{K}(k) - \omega^2 \mathbf{M} \right] \tilde{\bm{u}}_{\bm{k}} = 0 \label{eq:evp}, 
\end{align}
where $\mathbf{K}$ and $\mathbf{M}$ are the stiffness and mass matrix, respectively. This eigenvalue problem, Eq.~\eqref{eq:evp}, can be  solved by imposing the angular frequencies~$\{\omega_i, \mid i = 1, \ldots, N_{\omega} \}$ and solving for the wave vector components~$k$ and the corresponding \text{Bloch} eigenvectors~$\tilde{\bm{u}}_{\bm{k}_i}$. This $k(\omega)$-formulation allows for the characterization of both propagating and evanescent modes, as the imaginary part of the wave vector provides a measure of wave attenuation within the band gaps. The solutions can be obtained for any unit cell geometry using the finite element method~\cite{Hussein.2009}.
Here, we employed the Partial Differential Equations (PDEs) module of the commercial software \text{Comsol Multiphysics}~\cite{Comsol};  the implementation steps are detailed in Appendix~A of the supplementary material.
The solutions $\{k(\omega_i), \mid i = 1, \ldots, N_{\omega} \}$ are a discretised representation of the~\text{dispersion relation} plotted in a band structure diagram by limiting the real part of the wave vector to an irreducible part of the \text{Brillouin} zone, i.e.,~$\mathrm{Re}\{ k \} \in [0, \pi/a]$~\cite{Joannopoulos.2008, Deymier.2013}.
The eigenfrequencies can be normalized, e.g., with respect to the shear velocity $c_\mathrm{s} = \sqrt{G/\rho} = \sqrt{E/2\rho (1+\nu)}$ of the bulk material and the lattice period $a$ such that $f^* = f a/ c_\mathrm{s}$, for ease of comparison. 

\subsection{Modal characterization indices}
The modal characteristics of the \text{Bloch} eigenmodes~$\tilde{\bm{u}}_{\bm{k}}$, obtained from Eq.~\eqref{eq:evp}, can be analyzed by introducing modal indices. These indices can help identifying the dominant aspects of the wave motion and include (\textit{i}) an energy-equivalent polarization measure associated with dominant axial or transverse displacements, and (\textit{ii}) a rotational polarization measure that captures the rotational content and also distinguishes between axial and transverse contributions.

The first index quantifies the fraction of modal energy associated with the displacement along the periodic axis, i.e., the $z$-direction:
\begin{align}
    p_z(\omega, k) = \frac{\int_V |u_z|^2 \,\mathrm{d}V }{\int_V \left(|u_x|^2 + |u_y|^2 + |u_z|^2\right)\,\mathrm{d}V} \in [0,1],
    \label{eq:p_z}
\end{align}
where \( V \) denotes the unit cell volume. 
The second index captures the rotational content based on the curl of the displacement field, \( \bm{\psi} = \nabla \times \bm{u} \), and characterizes the axial rotational polarization of each mode as
\begin{align}
    p^{(\psi)}_z(\omega, k) = 
    \frac{\int_V |\psi_z|^2\, \mathrm{d}V}
         {\int_V \left(|\psi_x|^2 + |\psi_y|^2 + |\psi_z|^2 \right) \mathrm{d}V}
    \in [0,1].
    \label{eq:curl_polarization}
\end{align}
Averaging the rotational contributions in this manner permits the visual distinction of the lowest-frequency branches of the transverse modes, i.\,e.~shear and torsion, as well as the longitudinal modes. The modal characterization indices stated in Eq.~\eqref{eq:p_z} and Eq.~\eqref{eq:curl_polarization} are subsequently used to color-code the band structure and thereby distinguish between different dispersion branches.

\subsection{Analytical models for Bragg scattering and mode coupling}\label{sec:analyitcal}
\begin{figure}[]
\centering
\begin{subfigure}[t]{0.49\textwidth}
    \centering
    \includegraphics[width=.75\textwidth]{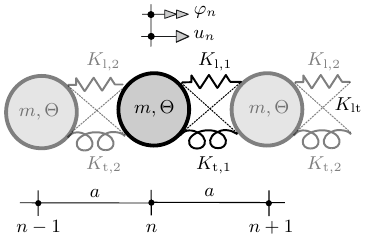}
    \caption{}\label{fig:MassSpring_LongiTorsion}
\end{subfigure}%
\hfill
\begin{subfigure}[t]{0.49\textwidth}
    \centering
    \includegraphics[width=.8\textwidth]{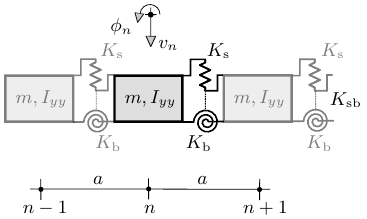}
\caption{}\label{fig:Analytic_flexural}
\end{subfigure}
\caption{Lumped mass--spring models for studying the low-frequency
dispersion and wave-interaction characteristics of the reference and twisted Kelvin cell chains. 
(a)~Longitudinal--torsional model with translational $u_n$ (along the propagation axis) and rotational $\varphi_n$
(about the propagation axis) degrees of freedom. Dashed lines indicate the linear elastic coupling between the degrees of freedom, with stiffness
$K_{\mathrm{lt}} = K_{\mathrm{tl}}$; the diatomic variant uses alternating stiffnesses
$K_{\mathrm{l},1}$ and $K_{\mathrm{l},2}$. 
(b)~Shear--bending model with transverse $v_n$  and rotational $\phi_n$ (about the axis normal to the shear direction) degrees of freedom. Dashed lines denote the coupling stiffness $K_{\mathrm{sb}} =
K_{\mathrm{bs}}$. }
\label{fig:AnalyticalModels}
\end{figure}

To complement the numerical Bloch-Floquet analysis and support the interpretation of calculated band-structure diagrams, a minimal analytical model is introduced that captures key wave-propagation features observed in twisted Kevin cell arrangements. 
The model builds on the initial contribution by~\text{Park}\,\textit{et al.}~\cite{Park.2022, Park.2022b} and represents a lumped-parameter mass-spring system, in which each unit cell has axial and torsional degrees of freedom, $u_n$ and $\phi_n$, representing the translational and rotational displacements of the unit cell, respectively (see Fig.~\ref{fig:MassSpring_LongiTorsion}). In contrast to classical uncoupled formulations, these degrees of freedom are linked through linear off-diagonal stiffness terms, which account for the geometric asymmetry introduced by twisting the Kelvin cell and thereby enable longitudinal–torsional coupled modes~(cf. Sec.~\ref{sec:Dispersion_Twisted}). 

The governing equations of motion for the unit $n$ read
\begin{align}
m \ddot{u}_n &= K_\mathrm{l} (u_{n+1} + u_{n-1} - 2u_n) + K_\mathrm{lt} (\theta_{n+1} - \theta_{n-1}), \label{eq:axial_eq} \\
\Theta \ddot{\varphi}_n &= K_\mathrm{t} (\varphi_{n+1} + \varphi_{n-1} - 2\varphi_n) + K_\mathrm{tl} (u_{n+1} - u_{n-1}), \label{eq:torsional_eq}
\end{align}
where $m$ is the effective mass, $\Theta$ the polar moment of inertia, $K_\mathrm{l}$ and $K_\mathrm{t}$ the longitudinal and torsional stiffness, respectively, and $K_\mathrm{lt} = K_\mathrm{tl}$ are the stiffness terms representing a symmetric longitudinal–torsional coupling, see Fig.~\ref{fig:MassSpring_LongiTorsion}. 

Assuming Bloch-type harmonic wave solutions of the form
\begin{align}
u_n(t) = \hat{u} e^{i(qna - \omega t)}, \quad \varphi_n(t) = \hat{\varphi} e^{i(qna - \omega t)},
\label{eq:BlochAnsatz}
\end{align}
Eqs.~\eqref{eq:axial_eq}–\eqref{eq:torsional_eq} reduce to the eigenvalue problem
\begin{equation}
\begin{bmatrix}
- m \omega^2 + 2K_\mathrm{l}(1 - \cos qa) & -2i K_\mathrm{lt} \sin qa \\
2i K_\mathrm{lt} \sin qa & - \Theta \omega^2 + 2 K_\mathrm{t}(1 - \cos qa)
\end{bmatrix}
\begin{bmatrix}
\hat{u} \\
\hat{\varphi}
\end{bmatrix}
= \bm{0}, 
\label{eq:disp_matrix}
\end{equation}
where $q$ denotes the scalar wave number of the mass-spring chain, corresponding to the \text{Bloch} wave vector~$\bm{k}$ introduced in Eq.~\eqref{eq:Bloch}.
The dispersion relation is obtained from the vanishing determinant of the coefficient matrix such that
\begin{equation}
\left(- m \omega^2 + 2K_\mathrm{l}(1 - \cos qa)\right) \left(- \Theta \omega^2 + 2K_\mathrm{t}(1 - \cos qa)\right) + 4K_\mathrm{lt}^2 \sin^2 qa = 0.
\label{eq:disp_relation}
\end{equation}
Although the Bragg-type band gap and longitudinal–torsional coupling dominate the attenuation behavior in the low frequency range, the twisted lattice also supports transverse -- flexural and shear -- wave modes. To capture these modes, flexural motion in a periodic chain is modeled following the approaches of~\text{Oh}\,\textit{et al.}~\cite{Oh.2016, Oh.2017}, where each unit cell is assigned the transverse displacement $v_n$ and the rotational angle $\phi_n$ about the axis perpendicular to the propagation direction. Using this formulation, transverse shear deformation is implicitly accounted for through the coupling between transverse displacement and rotational degrees of freedom, thereby jointly representing flexural and shear effects. Accordingly, the coupled equations of motion are given by
\begin{align}
m \ddot{v}_n &= K_\mathrm{s} \left( (v_{n+1} + v_{n-1} - 2v_n) + K_\mathrm{sb} \left(\phi_{n-1} - \phi_{n+1} \right) \right), \\
I \ddot{\phi}_n &= K_\mathrm{b} (\phi_{n+1} + \phi_{n-1} - 2\phi_n) + K_\mathrm{bs} K_\mathrm{s}(v_{n+1} - v_{n-1}) - K_\mathrm{bs}^2 K_\mathrm{s}(\phi_{n+1} + 2\phi_n + \phi_{n-1}),
\end{align}
where $K_\mathrm{s}$ is the vertical shear stiffness, $K_\mathrm{b}$ the bending stiffness associated with relative rotations between adjacent nodes, $K_\mathrm{bs} = K_\mathrm{bs}$ the shear–rotation coupling stiffness, and $I$ the inertia associated with shear rotation, see Fig.~\ref{fig:Analytic_flexural}.
These governing equations are analogous to those from the \text{Timoshenko} beam theory, in which shear deformation and rotational inertia are coupled through a shear–rotation interaction term~\cite{Park.2022b}.

The assumption on \text{Bloch}-type solutions $v_n = \hat{v} e^{i(qna - \omega t)}$, $\phi_n = \hat{\phi} e^{i(qna - \omega t)}$ leads to the following system:
\begin{equation}
\begin{bmatrix}
- m \omega^2 + 2K_\mathrm{s}(\cos qa -1) & 2i K_\mathrm{bs} \sin qa \\
-2i K_\mathrm{bs} \sin qa & - I \omega^2 + 2K_\mathrm{b}(\cos qa - 1) - 2K_\mathrm{s} K_\mathrm{bs}^2 (\cos qa + 1)
\end{bmatrix}
\begin{bmatrix}
\hat{v} \\
\hat{\phi}
\end{bmatrix}
= \bm{0}.
\label{eq:flexural}
\end{equation}
The corresponding solutions yield two coupled flexural branches that exhibit typical characteristics, such as band curvature, depending on $K_\mathrm{b}$ and $I$. Although these branches are not involved in the band gap formation discussed in this work, they appear in the numerical band structures and may contribute to broadband excitations or off-axis loading. Moreover, the analytical expressions in Eq.~\eqref{eq:disp_relation} allow for the estimation of the location and width of the band gap from a small set of effective parameters, which can be obtained from unit cell homogenization or numerical fitting. 

\subsubsection*{Mode similarity index for coupled modes}
Introducing linear-elastic stiffness coupling, as shown in Eq.~\eqref{eq:disp_matrix}, links formerly independent degrees-of-freedom and may thereby affect the purity of modal polarizations. In particular, off-diagonal stiffness terms may give rise to mixed polarizations, which can be identified from the associated eigenvectors. To consider the effect of coupling on the modal characteristics, we employ a \textit{mode similarity} index inspired by the modal assurance criterion~\cite{Allemang.2003}. This metric provides a basis for characterizing mode interaction beyond standard polarization color-coding by quantifying the geometric similarity between eigenvectors in avoided-crossing regions. Specifically, the similarity between eigenmodes is evaluated by computing an inner product matrix
\begin{align}
    \mathcal{O}^{(k)} 
    = \big| \left( \mathbf{\tilde{U}}^{(k)} \right)^H \mathbf{\tilde{U}}^{(k)} \big|, 
    \label{eq:innerProductMatrix}
\end{align}
where $\mathbf{\tilde{U}}^{(k)} = \left[ \mathbf{\tilde{u}}_1^{(k)}, \mathbf{\tilde{u}}_2^{(k)}, \ldots, \mathbf{\tilde{u}}_m^{(k)} \right] \in \mathbb{C}^{m\times m}$ collects \text{Bloch} eigenvectors obtained at wavenumber~$k$, $(\cdot)^H$ denotes the \text{Hermitian} transpose, and $m$ are the degrees-of-freedom considered. The individual entries of the inner product matrix, computed from the Euclidean inner product 
 \begin{align}
    \mathcal{O}_{jl}^{(k)} = \big| \left( \mathbf{\tilde{u}}_j^{(k)} \right)^H \cdot \mathbf{\tilde{u}}_l^{(k)} \big|,
 \end{align}
quantify the degree of geometric similarity between modes~$j$ and~$l$. Based on these entries, a mode similarity index~$\chi$ is defined as
\begin{align}
    \chi_j^{(k)} = \sum_{\substack{l=1 \\ l \neq j}}^{m} 
\big| \big( \mathbf{\tilde{u}}_j^{(k)} \big)^{H} \cdot \mathbf{\tilde{u}}_l^{(k)} \big|, 
\end{align}
which discards the self-similarity and sums the off-diagonal elements of mode~$j$ in $\mathcal{O}^{(k)}$. Large values of $\chi_j^{(k)}$ indicate strong similarity of the associated eigenvectors and thus modal interaction as typically observed in avoided crossing regions of dispersion relations~\cite{Mace.2012}. For visualization, the mode similarity index is normalized to the interval $[0, 1]$ such that
\begin{align}
    \tilde{\chi}_j^{(k)} = \frac{ \chi_j^{(k)} - \chi_\mathrm{min}}{\chi_\mathrm{max}- \chi_\mathrm{min}} \, \in [0,1], \;\;\chi_\mathrm{min} = \min_{j,k}\chi_j^{(k)}, \chi_\mathrm{max} = \max_{j,k}\chi_j^{(k)}.
    \label{eq:ModeCouplingIndex}
\end{align}
This normalized index is employed in Sec.~\ref{sec:Dispersion_Twisted} to color-code the dispersion relation obtained from the analytical model.

\section{Dispersion characteristics and band-gap formation through twisting}\label{sec:Dispersion_Discussion}

\subsection{Dispersion relation of the reference Kelvin cell}
\begin{figure}[]
    \centering
        \begin{subfigure}[t]{0.35\textwidth}
        \centering
        \includegraphics[width= .89\textwidth]{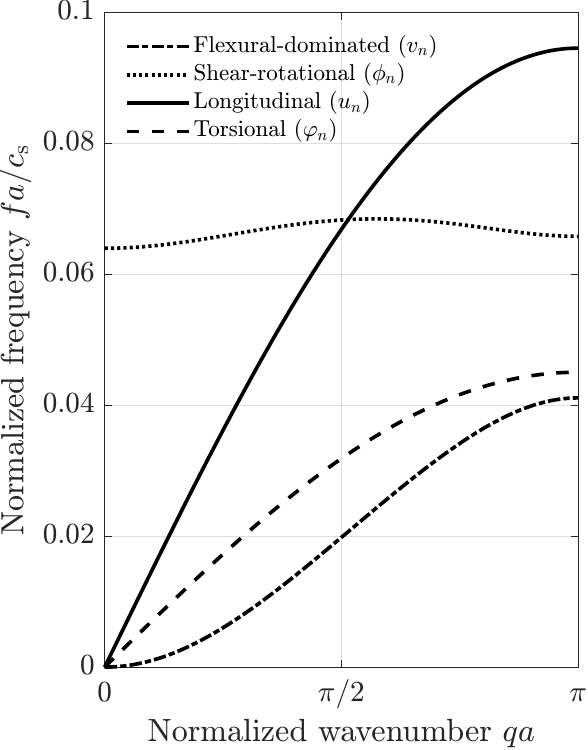}
        \caption{}\label{fig:Analytic_Dispersion_Isometric}
    \end{subfigure}    
    \hfill
   \begin{subfigure}[t]{0.63\textwidth}
        \centering
        \includegraphics[width=\textwidth]{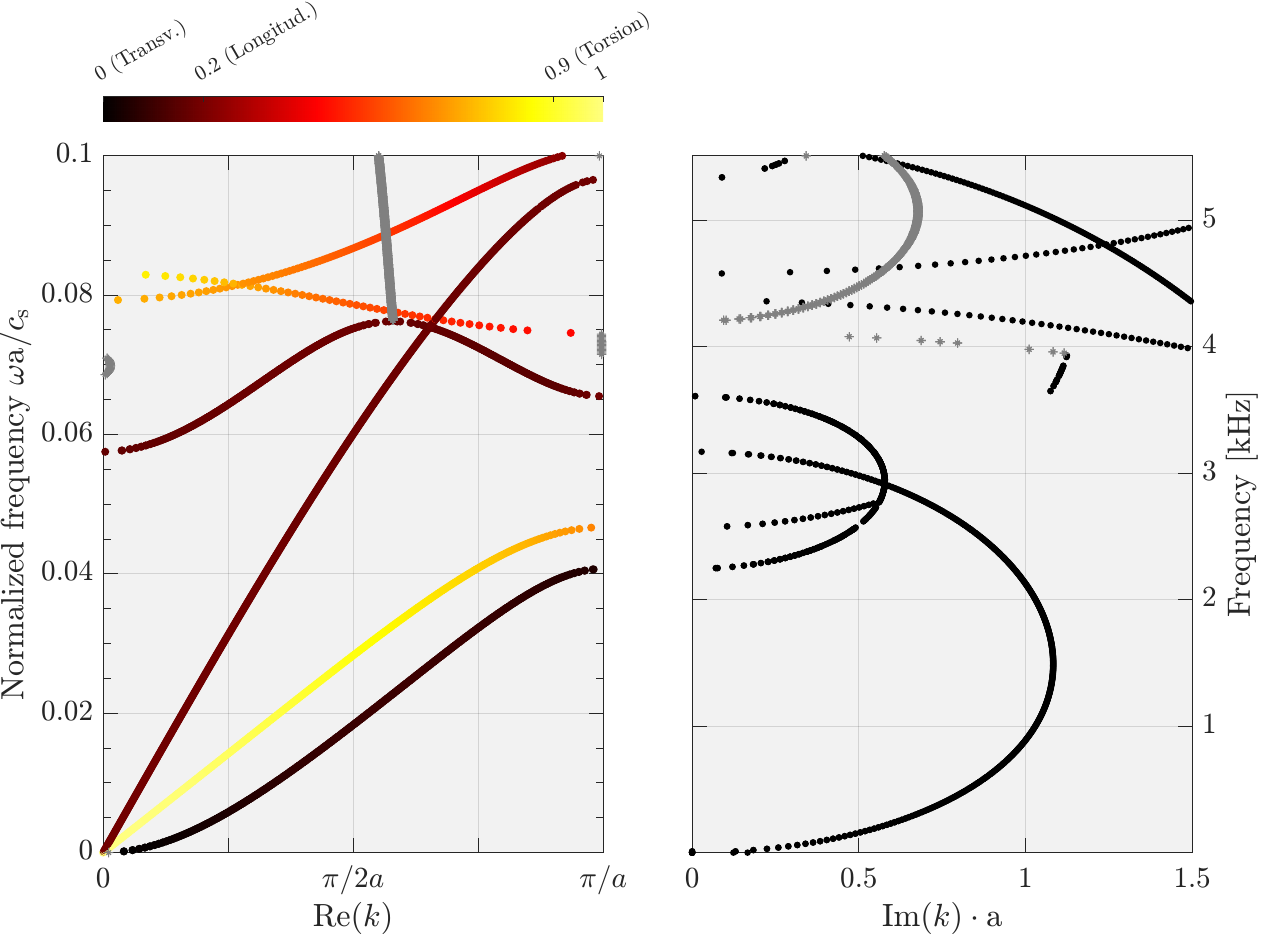}
        \caption{}\label{fig:KC_Dispersion}
        \end{subfigure} 
    \caption{Dispersion band structure of the reference Kelvin-cell chain and the corresponding analytical model.
    (a) Dispersion relations predicted by the analytical models, Eqs.~\eqref{eq:disp_relation} and~\eqref{eq:flexural}, reproducing the principal features of the low-frequency behavior of the Kelvin-cell chain, including the relative spacing of the longitudinal, torsional, and flexural branches. Model parameters were adjusted heuristically for overall correspondence and are listed in the supplementary material (Appendix~B).
    (b) Numerically estimated dispersion relation of the reference Kelvin-cell chain. Color-coding highlights the axial rotational polarization computed from Eq.~\eqref{eq:curl_polarization}, to identify the mode polarization. Evanescent modes with $\mathrm{Re}(k)=0$ are shown as black dots, while modes with $\mathrm{Re}(k)\neq0$, $\mathrm{Im}(k)\neq0$ are shown as gray squares.}\label{fig:KC_combined}
\end{figure}
Figure~\ref{fig:KC_Dispersion} presents the dispersion relation of the reference \text{Kelvin} cell, shown in Fig.~\ref{fig:2_Elementary_KC}, subjected to axial periodic translation. The unit cell is discretized using quadratic tetrahedral elements, resulting in a mesh of approximately $2.2\cdot10^4$~elements. This resolution was verified to ensure convergence of the eigenfrequencies in the considered frequency range. The material is assumed to be homogeneous, isotropic, and linear-elastic. We use $E = 4.1\,\mathrm{GPa}$, $\rho = 1250\,\mathrm{kg\cdot m^{-3}}$ and $ \nu = 0.35$, corresponding to the polymer in the additively manufactured specimens (Sec.~\ref{sec:Fabrication}).  Numerical results are complemented by the analytical dispersion relations in Fig.~\ref{fig:Analytic_Dispersion_Isometric}, which result from the coupled monoatomic mass-spring models introduced in Section~\ref{sec:analyitcal}.

In the analyzed frequency range, three fundamental branches originating from zero frequency dominate the band structure. These branches are properly captured by the analytical model (cf. Fig.~\ref{fig:Analytic_Dispersion_Isometric} and ~\ref{fig:KC_Dispersion}), as they 
correspond to the modes of a continuous, homogeneous, linear-elastic waveguide in the long-wavelength limit ($k \rightarrow 0$). Alternatively, they can be considered as the flexural, torsional, and longitudinal modes of the mono-atomic mass-spring chain discussed earlier. The first branch corresponds to \textit{shear} motion, involving dominant deflections in the $xy$-plane, perpendicular to the wave propagation direction. The second branch captures \textit{torsional} motion, characterized by rotational deformation about the propagation axis. The third branch, depicted in red, exhibits dominant \textit{longitudinal} polarization, associated with axial compression and extension along the propagation direction. At higher frequencies, the dispersion curves exhibit nonlinear characteristics, with mode shapes increasingly influenced by shear deformation of the lattice struts. Although complete band gaps do not appear in the frequency range considered, there is a region near $f^* = 0.05$ where only longitudinally polarized modes are supported by the infinite periodic structure. Therefore, one can define it as a band gap for flexural and torsional modes. This observation may be exploited further to induce a full band gap by proper adjustment of the microstructure. 

\subsection{Twisted configuration and band-gap mechanisms}\label{sec:Dispersion_Twisted}
\begin{figure}[]
    \centering
    \includegraphics[width = 0.5\textwidth]{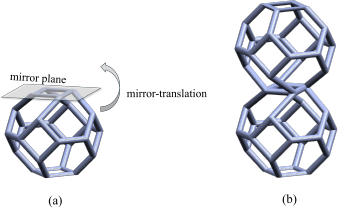}
    \caption{The design procedure of the twisted unit cell. (a) The Kelvin cell with a top face twisted by~$\theta_{z{+}}^{\hat{z}} = 45^\circ$ about the vertical axis. As this operation creates non-matching cell faces, the cell is duplicated by a mirror-translation with respect to the shown horizontal plane. (b) Resulting unit cell with a doubled characteristic length used in the subsequent analysis. }
    \label{fig:UnitCell_Twist}
\end{figure}

\begin{figure}[]
    \centering
    \begin{subfigure}[t]{0.315\textwidth}
        \centering
        \includegraphics[width= \linewidth]{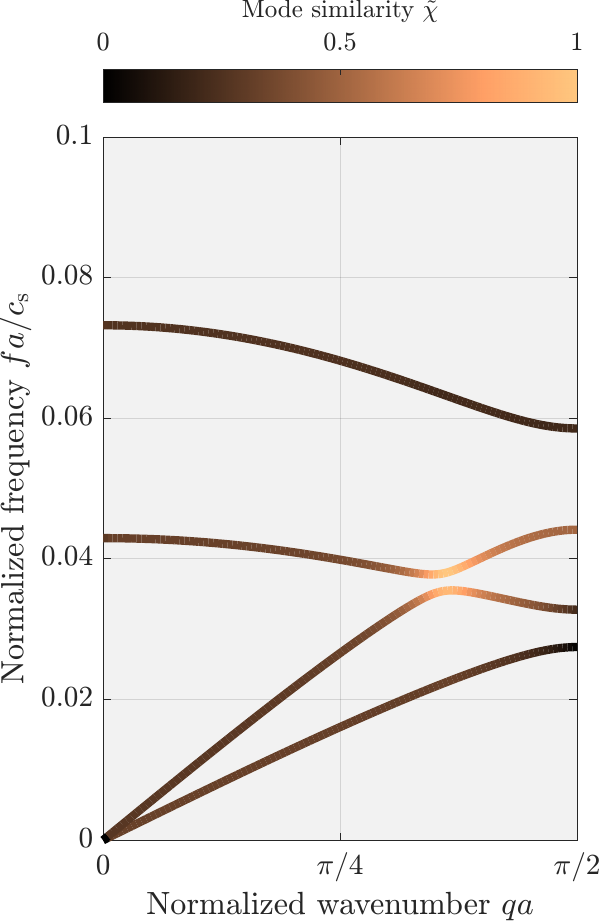}
        \caption{}\label{fig:3_Analytic_avoidCross}
    \end{subfigure}
    \hfill
    \begin{subfigure}[t]{0.65\textwidth}
        \centering
        \includegraphics[width=\linewidth]{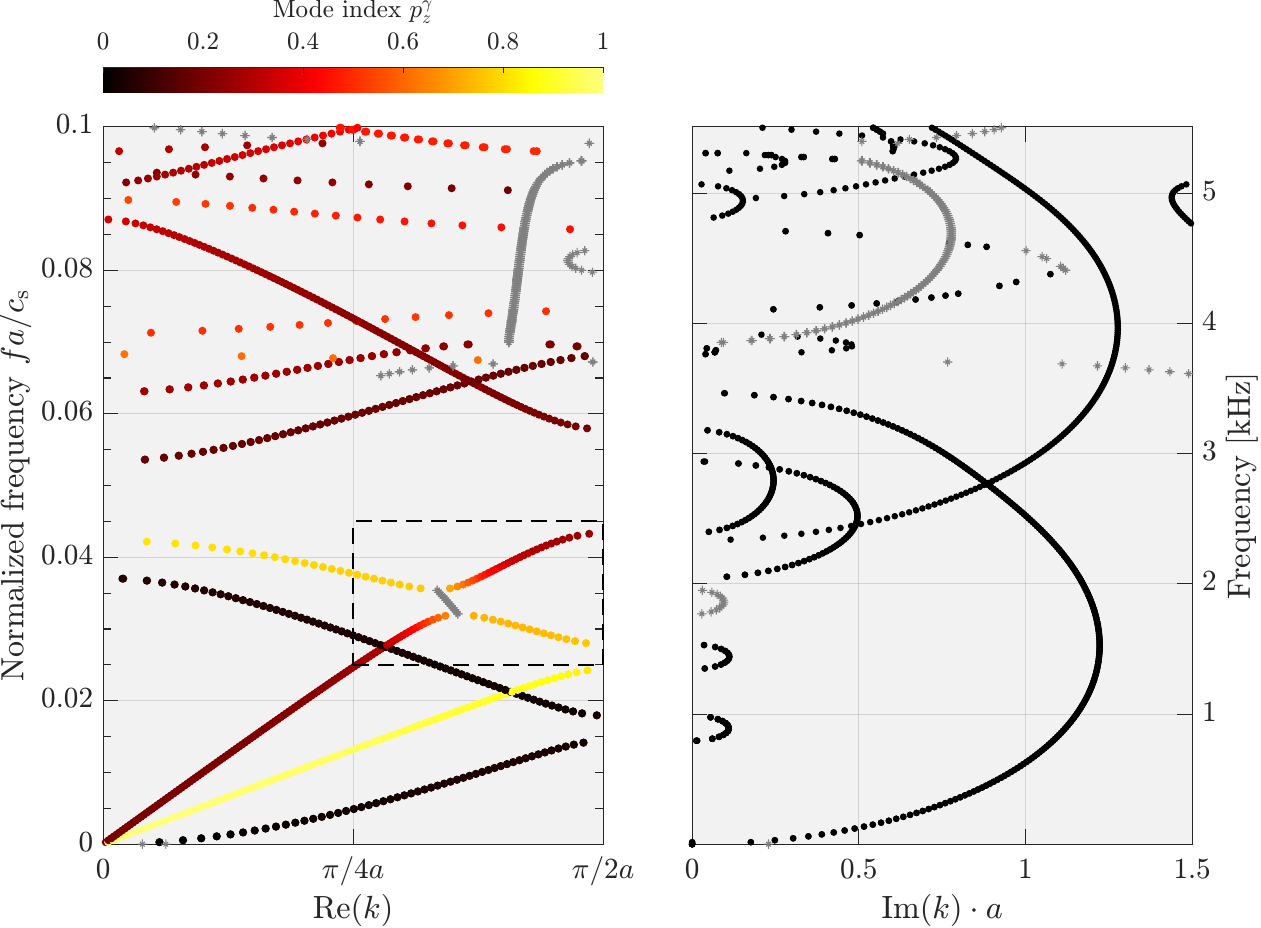}
        \caption{}\label{fig:Twist_Dispersion}
    \end{subfigure}
    \begin{subfigure}[b]{0.5\textwidth}
        \vspace{1.5em} 
        \centering
        \includegraphics[width=.85\textwidth]{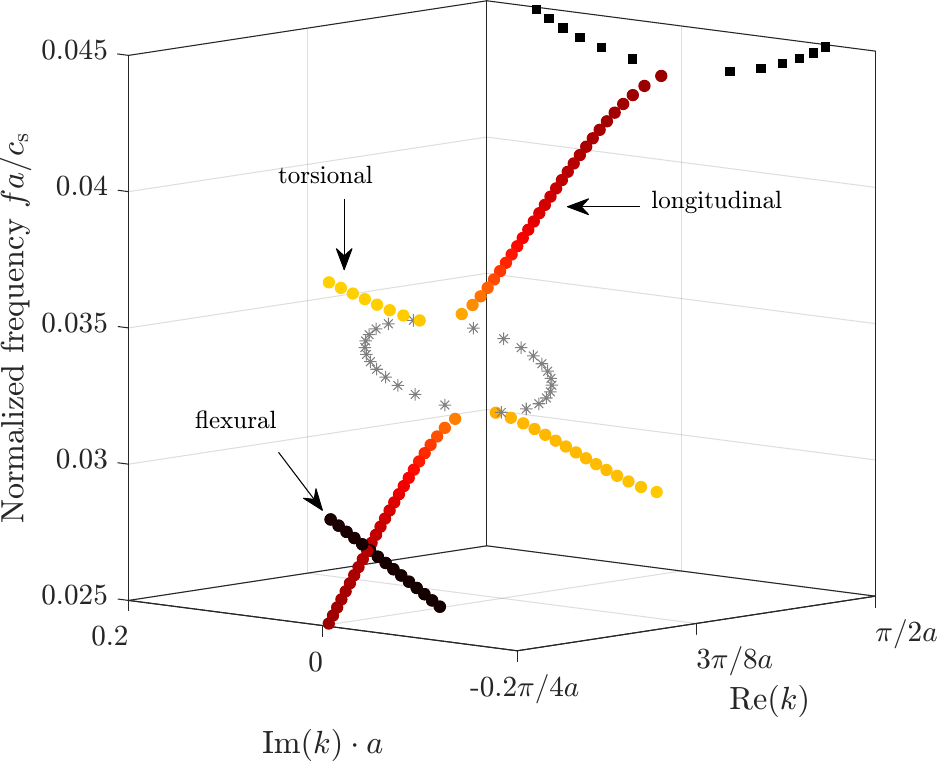}
        \caption{} \label{fig:Locking}
    \end{subfigure}
    \caption{Dispersion band structure of the twisted Kelvin-cell chain and the induced longitudinal–torsional coupling. 
    (a) The analytically derived dispersion relation of the coupled longitudinal–torsional model, highlighting mode interactions near $qa \approx \nicefrac{\pi}{3}$, as shown by the mode similarity index~$\tilde{\chi}$ (see Eq.~\eqref{eq:ModeCouplingIndex}). The relevant geometric parameters are given in in the supplementary material (Appendix~B).
    (b) The numerically estimated band structure of the $45^\circ$-twisted Kelvin-cell chain formed by the unit cell shown in Fig.~\ref{fig:UnitCell_Twist}, with the band gap and avoided crossing of the longitudinal and torsional branches. The color indicates the degree of axial rotation (see Eq.~\eqref{eq:curl_polarization}). Attenuated modes ($\mathrm{Im}(k) \ne 0$ and $\mathrm{Re}(k) \ne 0$) are shown as gray stars, while evanescent modes ($\mathrm{Re}(k) = 0$) appear as black dots. The dashed box marks the avoided-crossing region. 
   (c) A zoomed view of the avoided-crossing region, illustrating the coupling between longitudinal and torsional branches and the formation of the avoided crossing in the complex-valued plane.}
\label{fig:Dispersion_Combined}
\end{figure}

The potential for manipulating the eigenvalue spacing of the Kelvin cell unit cell through microstructural modifications is demonstrated using a twisted configuration shown in Fig.~\ref{fig:UnitCell_Twist}, where a simple geometric alteration is introduced by twisting only the top face through an angle of $\theta_{z{+}}^{\hat{z}} = 45^\circ$. This disorientation breaks the mirror symmetry of the unit cell, rendering the top and bottom square faces non-superimposable. As a result, periodic tessellation by pure translation is no longer feasible. Instead, glide symmetry is employed by reflecting the twisted unit cell with respect to the transverse plane at the unit cell boundary. This operation connects the twisted and untwisted faces along the waveguide axis, forming a supercell with double translational periodicity $\hat{a} = 2a$. The twisted configuration thus encompasses both a geometric transformation at the unit-cell level and an adapted tessellation strategy.

Figure~\ref{fig:Twist_Dispersion} presents the corresponding band structure of the described twisted Kelvin cell unit. The dispersion relation appears single-fold within the first \text{Brillouin} zone due to the applied supercell approach. Two principal observations emerge:

First, the real part of the band structure diagram reveals a band gap between the 
longitudinal branches around the normalized frequency~$f^* \approx 0.05$. In this frequency range, only evanescent modes with a non-zero imaginary component exist, indicating the presence of a full band gap. Clearly, this band gap is opened due to twisting and glide-symmetric tessellation of the reference Kelvin cell (compare  Figs.~\ref{fig:KC_Dispersion} and \ref{fig:Twist_Dispersion}). The imaginary part of the band structure diagram within the gap provides insight into the underlying attenuation mechanism. The evanescent branch linking the two longitudinal modes across the band gap shows a continous variation of $\mathrm{Im}(k)$ with frequency, reaching a peak value $\max \mathrm{Im}(k)$ at the mid-gap frequency $f^* \approx 0.05$. This profile is characteristic of \text{Bragg}-type scattering, which arises from destructive interference when the wavelength is comparable to the lattice periodicity~\citep{Yuan.2013}. For longitudinal waves, the Bragg frequency is estimated as $f_\mathrm{Bragg} \approx c_\mathrm{p}/2a$, where $c_\mathrm{p}$ is the effective phase velocity extracted from the slope of the acoustic longitudinal branch near $k \rightarrow 0$, resulting in $c_\mathrm{p} \approx 221$~m/s. This leads to $f_\mathrm{Bragg} \approx 2766$~Hz, corresponding to $f^*_\mathrm{Bragg} = 0.05$, in good agreement with the mid-frequency of the band gap.

The second feature is observed near $f^* = 0.035$, where the interaction between the longitudinal and torsional modes results in mode coupling. This interaction can be further traced in Fig.~\ref{fig:Locking}, which shows the complex dispersion relation. The branches associated with longitudinal and torsional motions initially converge and then diverge, resulting in an avoided crossing. Each branch exhibits a turning point where the group velocity vanishes ($\partial \omega / \partial k = 0$), and the branches reverse slope, implying counter-propagating wave modes. Around these turning points, the polarization states mix, converging toward $p_z^\psi \approx 0.7$, which indicates that the modes involve combined longitudinal and torsional deflections. This coupling results in a narrow frequency range in which both longitudinal and torsional modes are attenuated. However, because flexural modes continue to propagate in this region, a complete band gap cannot form. Instead, it is a polarization-dependent band gap resulting from mode interaction. The two complex branches in this regime exhibit exponential decay or growth, with Im$(k) < 0$ or Im$(k) > 0$, respectively, corresponding to attenuation and amplification.

The analytical mass–spring model providing the dispersion relation~\eqref{eq:disp_matrix} shown in Fig.~\ref{fig:MassSpring_LongiTorsion} captures both the position of the Bragg-type band gap and the coupling between the longitudinal and torsional branches. This is enabled by extending the monoatomic chain to a diatomic configuration with alternating stiffnesses that can adequately capture the supercell arrangement of a twisted Kelvin-cell chain (see supplementary material, Appendix B). The model further reveals that twisting modifies the stiffness distribution within the unit cell, thereby altering the phase velocities and facilitating mode coupling. Meanwhile, the glide-symmetric tessellation mirrors the torsional branch within the Brillouin zone, creating conditions for its interaction with the longitudinal branch. The resulting coupling, quantified by the coupling coefficient~$K_\mathrm{lt}$, leads to an avoided crossing, confirming that the combined effect of twisting and glide symmetry drives the formation of the polarization band gap. This avoided-crossing region is further identified by elevated values of the mode similarity index~$\tilde{\chi}$, approaching unity near the zero group-velocity points, which indicates pronounced mixed polarization.

In contrast to classical locally resonant band gaps, which arise from interactions between propagating waves and subwavelength resonators~\citep{Iorio.2023}, the mechanism observed here is due to coupling between two counter-propagating modes enabled by symmetry properties of the twisted geometry. The mirror-reflection symmetry imposed by the twist effectively reverses the propagation direction of the torsional mode in the dispersion diagram, allowing interaction with the longitudinal mode. This interaction has been described in the literature using terms such as \textit{mode locking}~\citep{Mace.2012}, \textit{mode matching}~\citep{Perkins.1986, Giannini.2016} or \textit{avoided crossing}~\citep{Achaoui.2010}. Unlike locally resonant band-gap mechanisms, which require the introduction of additional masses, the twisted Kelvin cell demonstrates that carefully controlled geometry and breaking symmetry are sufficient to create and tune wave-attenuation features. This behavior is described by the analytical model and is, furthermore, consistent with previous theoretical treatments of mode coupling in waveguides, as seen, e.g., in~\text{Mace}~\citep{Mace.2012} and others.

\subsection{Effects of the twist angle on the band-gap tunability}
\begin{figure}[]
     \centering
     \begin{subfigure}[t]{0.49\textwidth}
         \centering
         \includegraphics[width=.9\textwidth]{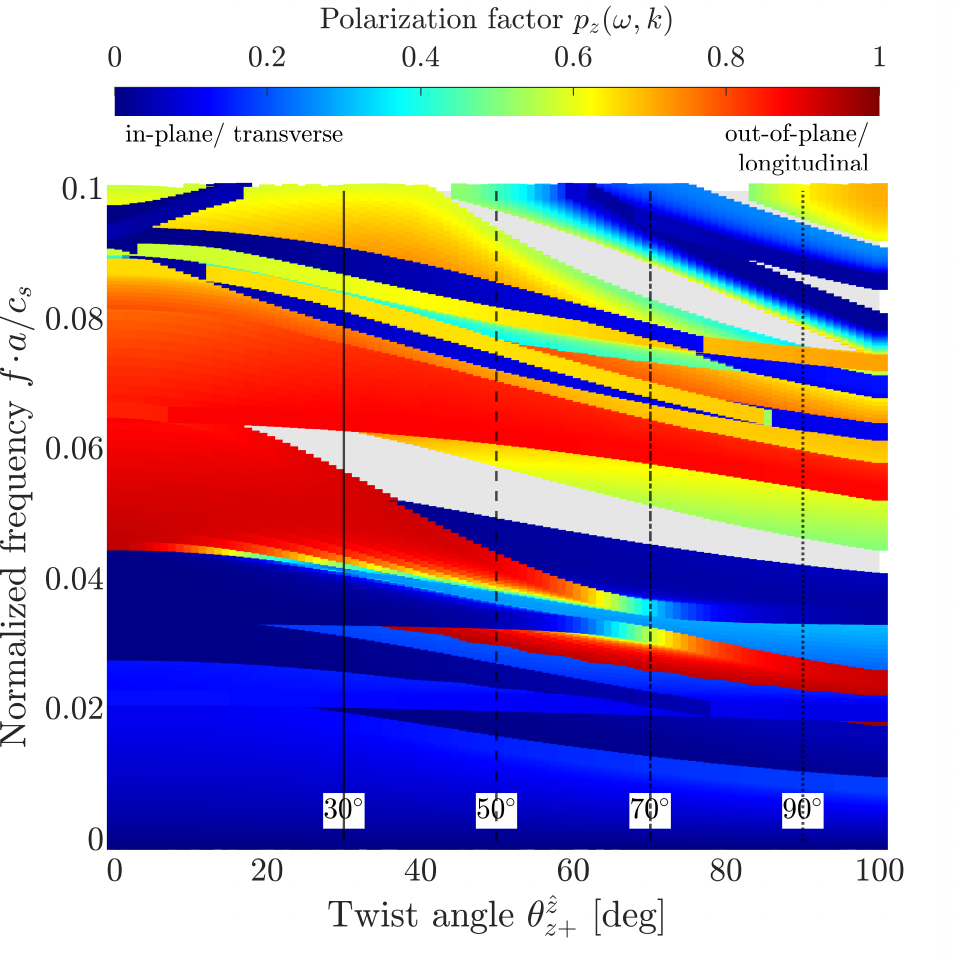}
         \caption{}\label{fig:Condensed}
     \end{subfigure}
     \hfill
     \begin{subfigure}[t]{0.49\textwidth}
         \centering
         \includegraphics[width=.9\textwidth]{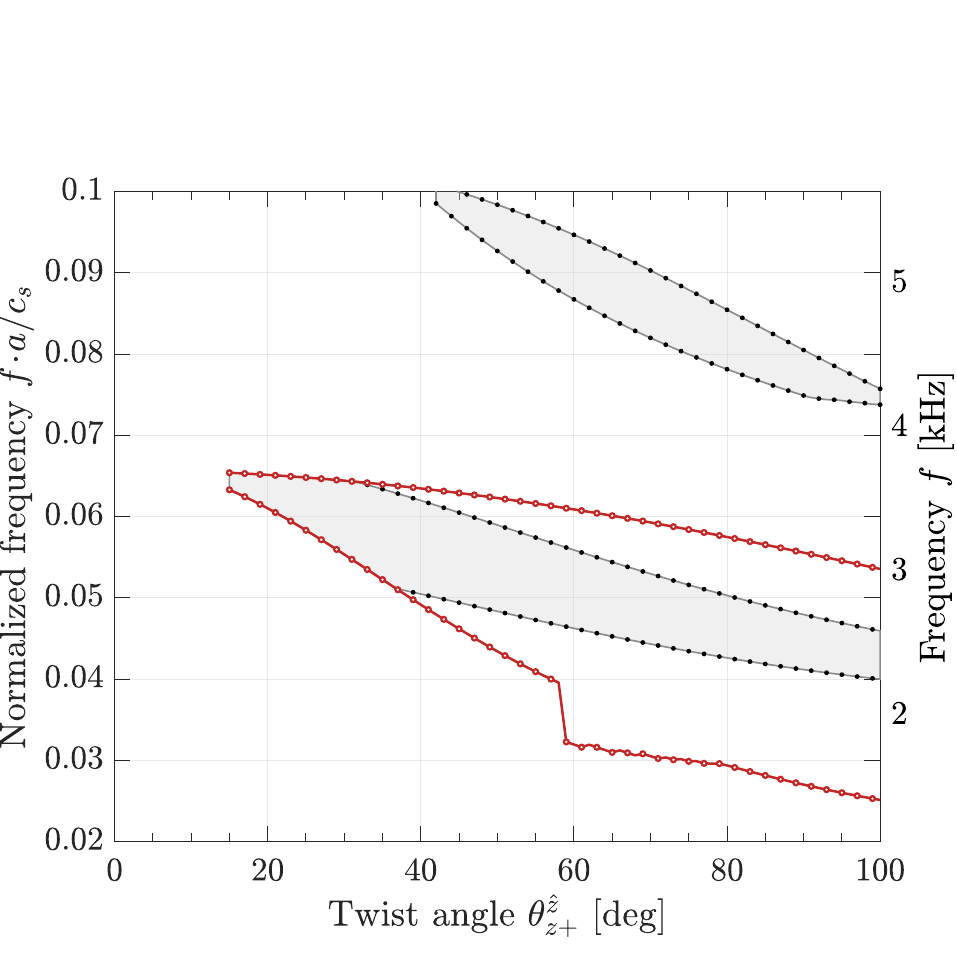}
         \caption{}\label{fig:Evolutuion}
     \end{subfigure} %
    \begin{subfigure}[]{\textwidth}
         \centering
         \includegraphics[width=\textwidth]{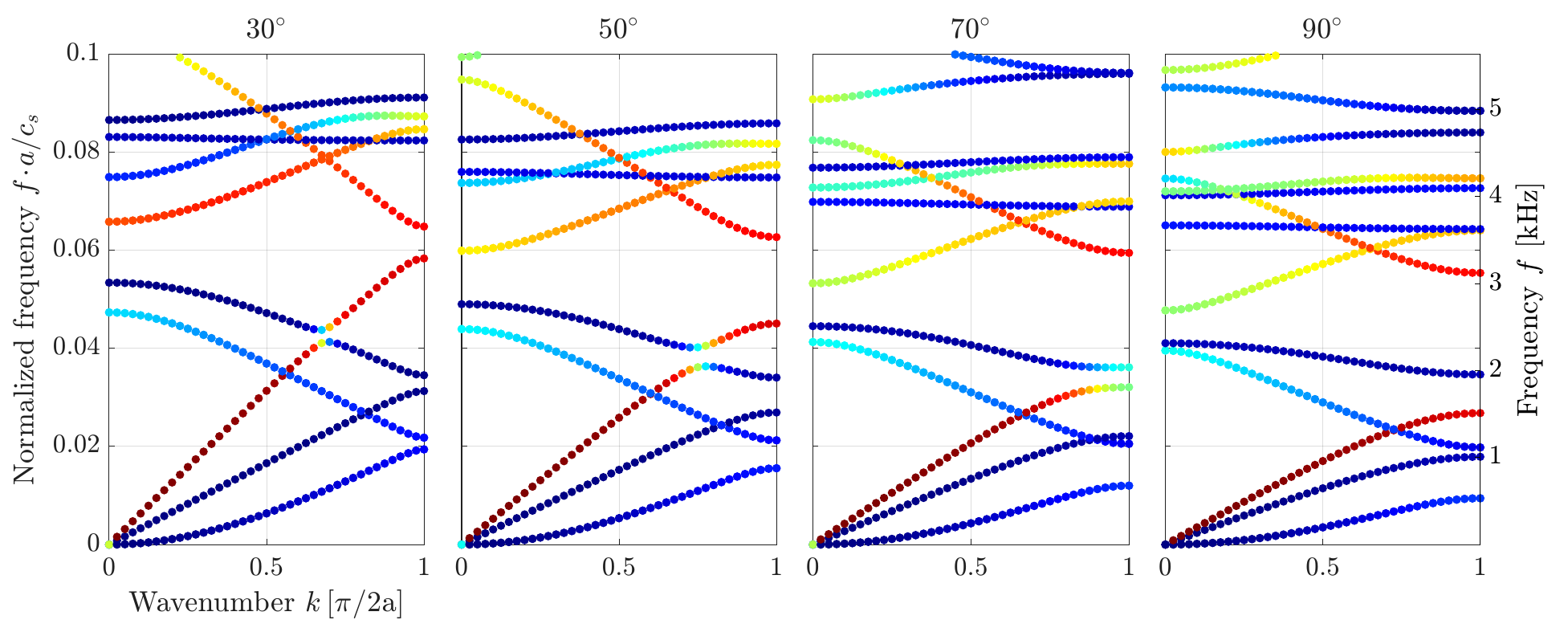}
         \caption{}\label{fig:Individual_Disp}
     \end{subfigure}
    \caption{Influence of the twist angle on the eigenfrequency spacing, band-gap tunability, and dispersion characteristics of the Kelvin-cell chain. 
    (a) The condensed dispersion diagram of the twisted Kelvin-cell chain, where the color denotes the dominant polarization factor~$p_z$ (see Eq.~\eqref{eq:p_z}), ranging from transverse/shear-dominated modes (blue) to axial/longitudinal-dominated modes (red). Gray regions mark complete band gaps. 
    (b) The evolution of the first two complete band-gap regimes (gray) and the frequency split between the folded longitudinal branches at the Brillouin zone edge $k=\pi/2a$ (red) as functions of the twist angle; black dots indicate the bounding frequencies of each band gap. 
    (c) Individual dispersion diagrams at twist angles corresponding to the vertical markers in~(a).}\label{fig:Dispersion_Parametric}
\end{figure}

To investigate the dependence of the complete \text{Bragg}-type band gap on the twist angle, a dispersion analysis is conducted to incrementally increase the twist angle~$\theta_{z{+}}^{\hat{z}}$ in steps of $1^\circ$, covering the range $[0^\circ, 100^\circ]$. To reduce computational cost, only the \textit{propagating} modes are computed using the standard formulation $\omega(k)$, often referred to as the \textit{inverse} approach~\cite{Claeys.2013, Cool.2024}. The calculated eigenfrequency distribution $\omega\left(k, \theta_{z{+}}^{\hat{z}} \right)$ is shown in Fig.~\ref{fig:Condensed} as a condensed visualization of all band structures in a single diagram. For each increment of the twist angle, the frequency values across the first irreducible \text{Brillouin} zone are projected onto a single vertical line, effectively collapsing the dimension $\operatorname{Re}(k)$ to highlight the evolution of band gaps with~$\theta_{z{+}}^{\hat{z}}$.

The condensed diagram reveals that increasing the twist angle causes a systematic shift of the propagating modes towards lower frequencies. This tuning effect is primarily driven by stiffness, as the mass contribution introduced by twisting is small compared with the extension and reorientation of the ligaments. At the maximum twist angle (100$^\circ$), the total mass increases by 9\;\% compared to the reference \text{Kelvin} cell.

The frequency shifts are mode-dependent and vary according to polarization as follows. "In-plane" polarized modes (shown in blue) exhibit minimal sensitivity, while longitudinally polarized modes (shown in red) are significantly affected. For the latter, the frequency gap between the acoustic and optical branches at the edge of the first \text{Brillouin} zone widens with increasing twist angle (compare with Fig.~\ref{fig:Evolutuion}), an effect reminiscent of branch separation in a diatomic mass–spring chain with diverging stiffness parameters.

The complete band gap (indicated in black in Fig.~\ref{fig:Evolutuion}) of a width greater than 100\,Hz emerges for the twist angles between $15^\circ$ and $100^\circ$. Increasing the twist angle lowers the band gap mid-frequency but does not increase the gap width, in contrast to the steady widening observed in the longitudinal branch separation. The maximum width of the band gap is achieved at a twist angle of $40^\circ$, corresponding to a relative width of $\nicefrac{(f_\mathrm{top} - f_\mathrm{bot})}{f_\mathrm{mid}} = 21.5\,\%$. For larger twist angles, the band gap narrows again due to the downward shift of "in-plane" polarized modes, which approach, and eventually bound the band gap.

The sudden decrease in the longitudinal edge mode frequency, observed in Fig.~\ref{fig:Evolutuion} around $60^\circ$ twist angle, occurs because the avoided crossing between the longitudinal and torsional branches disappears. As the twist angle increases, the avoided crossings and the points of zero group velocity between the longitudinal and torsional branches are gradually displaced towards higher wave numbers, approaching the edge of the first \text{Brillouin} zone, as illustrated in Figure~\ref{fig:Individual_Disp} for selected twist angles. Specifically, around $70^\circ$, coupled modes appear close to the edge of the Brillouin zone. However, for larger twist angles, the coupling conditions are no longer satisfied, and the two branches decouple so that the zero group velocity points inside the Brillouin zone vanish, as exemplified for $90^\circ$ in Fig.~\ref{fig:Individual_Disp}. 

To summarize, this parametric study demonstrates that the twist angle is an effective tuning parameter that not only controls the opening and closing of band gaps but also governs the occurrence of mode coupling and the location of zero-group-velocity points. Through this tunability, both the mid-frequency and the width of the band gap can be modified, providing an additional design lever for tailoring wave dispersion and energy localization in architected structures. However, the maximum attainable band-gap width is inherently limited and comes at the expense of reduced axial stiffness of the unit cell. Besides, progressive frequency splitting between longitudinal branches suggests the potential for achieving broader wave attenuation ranges in forced vibration scenarios, particularly those involving excitation along the periodicity dimension. These aspects are further explored in subsequent experimental investigations of finite-size structures.

%% file: content/04_Experiments.tex
\section{Experimental Validation}\label{sec:experiments}
To validate the numerical and analytical predictions, the wave attenuation functionality was experimentally tested on finite-size lattice structures in transmission (pitch-catch) tests. The frequency-dependent transmission of forced vibrations between two observation points within the lattice was measured to quantify the wave-filtering capabilities of the specimens.

\subsection{Samples manufacturing}\label{sec:Fabrication}
Based on the dispersion analysis in Fig.~\ref{fig:Evolutuion}, three lattice configurations were selected and manufactured for experimental testing: the untwisted reference Kelvin cell, a $45^\circ$-twisted specimen showing a wide \text{Bragg} band gap, and a $90^\circ$-twisted variant.

Each sample was formed by a $1\times1\times3$ array of twisted Kelvin cells, resulting in a total structure length of $L = 120$\,mm with a unit cell size of $a = 40$\,mm. These geometric parameters were selected to ensure compatibility with manufacturing constraints and maintain at least three unit cells along the axial direction. Each strut had a thickness of 1.5\,mm, resulting in a low lattice density~(Fig.~\ref{fig:Transmission_Experimental}). Square plates with a width of $w = 20$\,mm and a thickness of 1.5\,mm were attached at both ends of the sample to facilitate sensor mounting for vibration measurements. All specimens have the same overall length; however, the twisted variants contain three periodic supercells resulting from the glide-symmetric arrangement, whereas the regular Kelvin cell structure consists of six conventional translational unit cells. Illustrations of all tested configurations are provided in the supplementary material.

Of each configuration, two samples were made using stereolithography (SLA) on a commercial \text{Formlabs}3 SLA printer. We used \textit{Rigid4K} resin with a 3D-printed layer thickness of 50\,nm. 
According to the manufacturer's specifications, the cured resin has a \text{Young}'s modulus of $E = 4.1\,\mathrm{GPa}$, and a density of $\rho = 1250\,\mathrm{kg/m^3}$. A \text{Poisson} ratio of $\nu = 0.35$ was assumed, consistent with common values for polymers. 

After the 3D-printing process, the samples were ultrasonically cleaned in isopropanol to remove uncured resin, and then post-cured at 80\,$^\circ$C under 405\,nm light for 20 minutes using a \text{Formlabs} Cure oven. To avoid warping during curing, the temperature gradually increased from room temperature at a rate of 1\,$^\circ$C per minute. 

The open-cell architecture of the samples, combined with their low density, requires auxiliary support structures distributed throughout the geometry to prevent ligament deformation or collapse due to gravity.
The support structures were removed after post-curing to ensure dimensional stability. Finally, each sample was attached to a \text{10-32} UNF-threaded connector, which was 3D printed from the same resin. This step aims to minimize the impedance mismatch between the sample and the vibration excitation interface.

\subsection{Measurement setup}
\begin{figure}[]
    \centering
    \includegraphics[width = 0.9\textwidth]{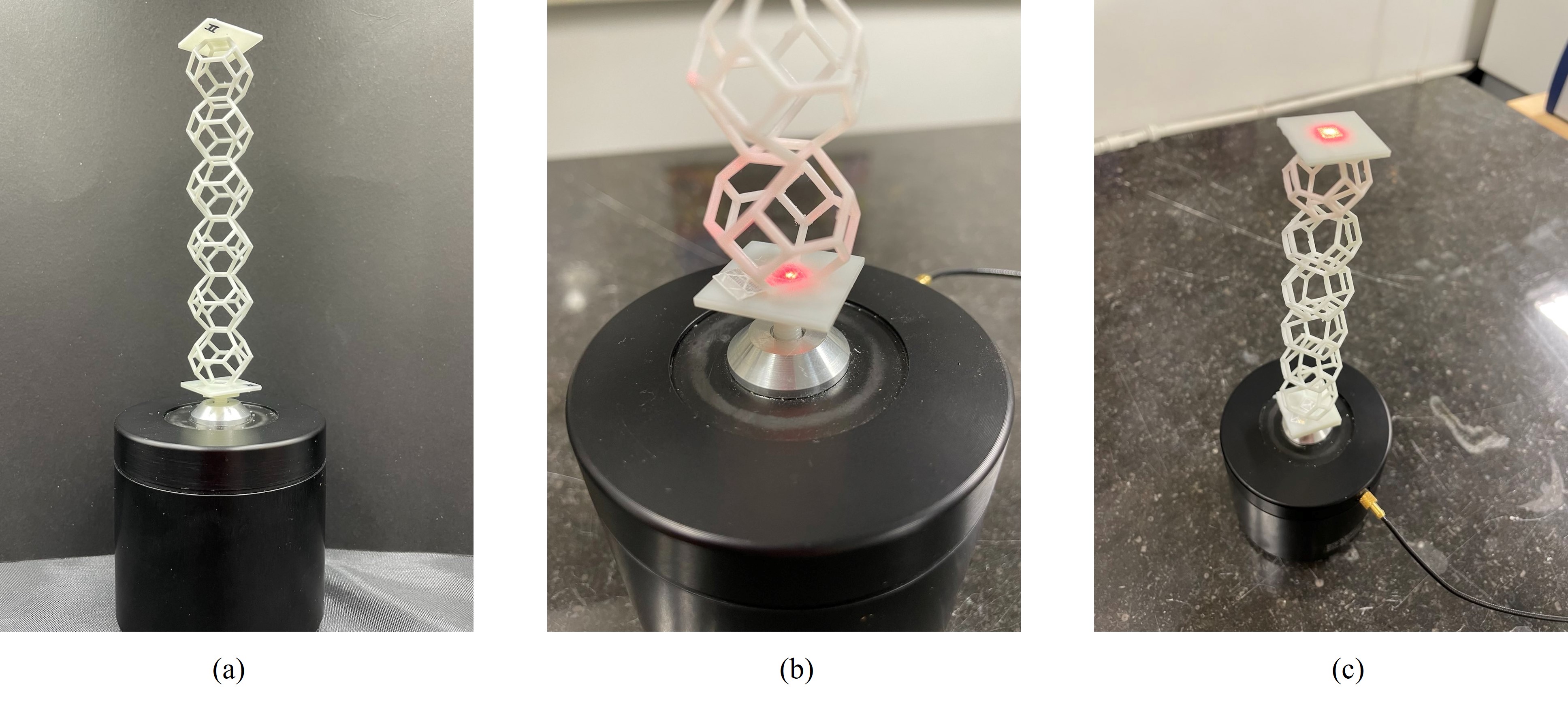}
    \caption{Setup for the transmission tests. (a)~Vertical alignment of the sample comprising three unit cells on the shaker; (b-c) Excitation and acquisition points at the bottom and top plates, respectively. The shown specimens are obtained as a three-times replication of the 45$^\circ$-twisted unit cell shown in~Fig.~\ref{fig:UnitCell_Twist}.}\label{fig:Setup}
\end{figure}
Each sample was vertically mounted using the UNF-connector onto the vibration shaker (Brüel \& Kjær Type 4810) driven by an amplifier (Type 2718), as illustrated in Fig.~\ref{fig:Setup}. The axial input and output velocity amplitudes, aligned with the direction of wave propagation, were measured at the bottom and top end plates, respectively, by means of a \text{Polytec}~OFV-5000 laser Doppler vibrometer. At each end plate, two measurement positions -- central and eccentric -- were selected to measure the vertical velocity component. For this, reflective tape was applied to the plates' surfaces at each measurement point to enhance the signal-to-noise ratio.

The excitation sources were two consecutive linear frequency sweeps: a low-frequency sweep from 100\,Hz to 1\,kHz and a high-frequency sweep from 1\,kHz to 10\,kHz, each lasting 4 seconds. Signal generation, synchronization, and data acquisition were performed using a \text{Spectrum} hybridNETBOX system at a 128\,kHz sampling rate. For the high-frequency sweep, the input signal amplifier's voltage was increased to improve sensitivity without exceeding its linear operating range, thereby avoiding signal distortion and maximizing measurement fidelity, particularly in the expected band-gap frequency ranges.

Post-processing of the acquired time-domain data was performed using in-house code in \text{MATLAB}. It included time-domain filtering, \text{Fourier} transformation, normalization, and spectral analysis to extract transmission characteristics.

\subsection{Experimental results}
\begin{figure}[]
    \centering
    \includegraphics[width = 0.6\textwidth]{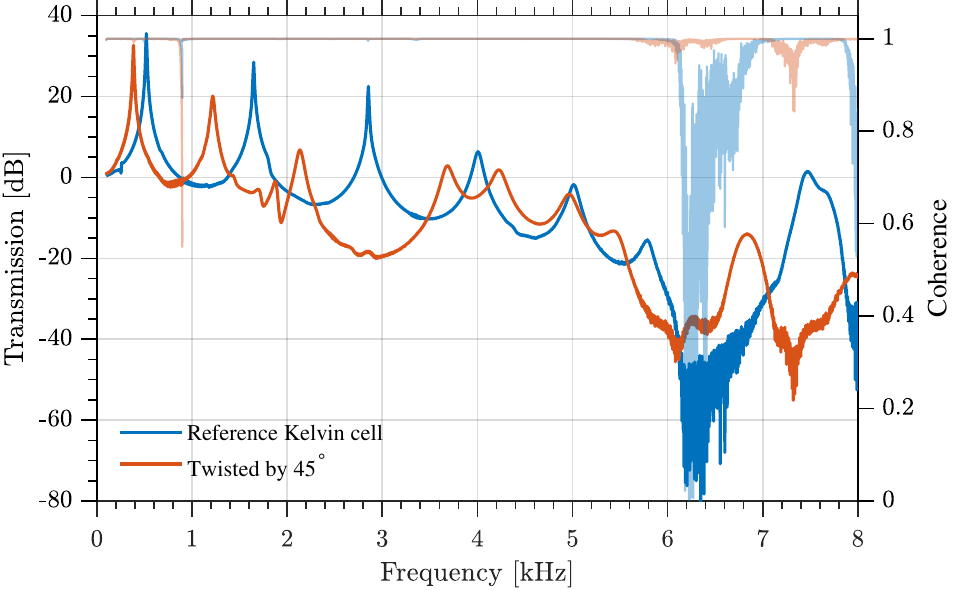}
    \caption{Measured transmission spectra (left axis, solid lines) and the corresponding coherence (right axis, faded lines in matching colors) for the SLA-printed reference and $45^\circ$-twisted Kelvin-cell (Fig.~\ref{fig:Setup}) chains composed of three unit cells. The coherence was computed from the
    measurements at the excitation and acquisition points. Both tested structures have identical external dimensions but differ in their underlying spatial periodicity pattern.}\label{fig:Transmission_Experimental}
\end{figure}

The measured transmission spectra of both the $45^\circ$ twisted configuration and the reference Kelvin cell structure are shown in Fig.~\ref{fig:Transmission_Experimental} by the curves in bold. The light shaded curves show the magnitude-squared signal coherence between the input and output velocity signals, computed following \text{Welch’s} averaged, modified periodogram method~\cite{Welch.1967}
as implemented in \text{Matlab's} \textit{mscohere} function. At frequencies below 6~kHz, coherence values close to unity indicate a linear relationship between the input and output responses, suggesting low noise and supporting the reliability of the measured data. At frequencies above 8~kHz, the coherence functions drop to around 0.5, indicating substantial noise and reduced signal reliability; therefore, the transmission data at these frequencies are excluded from consideration. A noticeable drop in coherence at~$\SI{0.9}{\kilo\hertz}$ marks the point of fusion for two separate measurement sweeps (low- and high-frequency excitation), which were combined to form the full transmission spectrum.

The effects of twist can be understood by comparing the transmission spectra of the two configurations, specifically in the wave-attenuation zones.
Around 2.9~kHz, the reference Kelvin cell structure exhibits an amplification peak of 20~dB. In contrast, the twisted arrangement reveals \text{Bragg} scattering at the same frequencies, leading to a reduction of up to 20~dB in the input signal. Therefore, the twist of the Kelvin cell can effectively alter the vibrational response and thus reduce wave transmission at target frequencies.

Further analysis of the transmission data is focused on the twisted configuration. The transmission spectrum reveals three prominent attenuation bands centered on approximately~$\SI{3}{\kilo\hertz}$, $\SI{6}{\kilo\hertz}$, and $\SI{7.3}{\kilo\hertz}$. In each of these regions, the transmission magnitude drops significantly, accompanied by coherence dips at the higher-frequency attenuation bands. This indicates strong wave attenuation in the lattice structure, which is consistent with the predicted band gap frequencies. The coherence dips on these frequencies further highlight the effectiveness of the structure in filtering vibrational energy.

\begin{figure}[]
    \centering
    \includegraphics[width = 0.7\textwidth]{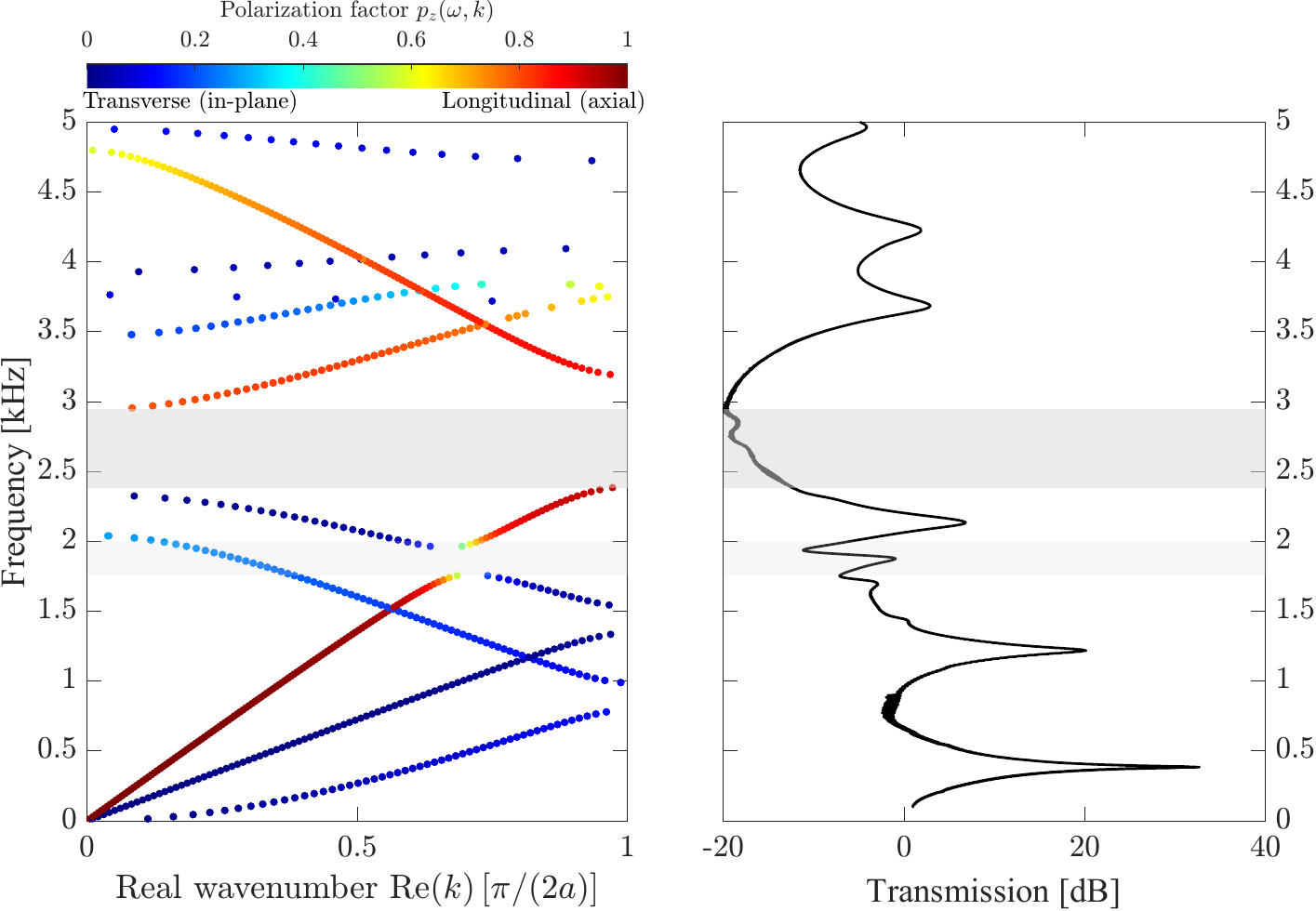}
    \caption{Dispersion diagram for the $45^\circ$-twisted Kelvin cell lattice \textit{vs.} measured wave transmission on a three-cell specimen.
    The color in the diagram indicates mode polarization~$p_z$ (see Eq.~\eqref{eq:p_z}), with red corresponding to longitudinal (axial) modes and blue to transverse (shear-dominated) and torsional modes. The experimental excitation was applied along the axial direction (see Fig.~\ref{fig:Setup}), predominantly activating the longitudinal modes. The dark-gray regions mark complete band gaps, while the light-gray region highlights the avoided crossing between the longitudinal and torsional branches.}
    \label{fig:BandLayout}
\end{figure}
Fig.~\ref{fig:BandLayout} shows the calculated band structure diagram for an infinitely periodic lattice with the measured transmission spectrum of the SLA structure, for frequencies up to~$\SI{5}{\kilo\hertz}$. The comparison demonstrates a strong correlation between theoretical \text{Bragg}-type (around~$\SI{3}{\kilo\hertz}$) and coupling-induced (around~$\SI{2}{\kilo\hertz}$) band gaps and the experimentally observed attenuation frequency ranges.
In the coupling-dominated frequency range, the transmission spectrum reveals a resonance–like pattern, with a peak attenuation of approximately \SI{11}{\decibel}. This indicates that the coupling mechanism, while effective, induces less pronounced attenuation than the \text{Bragg}-type band gaps. In contrast, the \text{Bragg}-induced attenuation frequencies, associated with the splitting of longitudinal branches in the dispersion diagram, reveal strong wave suppression, with maximum attenuation reaching \SI{20}{\decibel}, analogous to that for acoustic-optical branch separation in a diatomic mass–spring chain. 

The experimental findings confirm that the twisting-induced band gaps identified through dispersion analysis are reproduced in finite-size samples, validating the underlying model and underscoring the value of the twisting operation for enhanced vibration control.  Moreover, we demonstrated that even a few \text{Kelvin} unit cells are sufficient to achieve significant wave attenuation.

\subsection{Numerical vs. Experimental Transmission Analysis}
Despite the general agreement between the frequencies of the predicted band gaps and transmission dips in Fig.~\ref{fig:BandLayout}, some quantitative discrepancies are found. They originate mainly due to two factors: \textit{(i)} the comparison involves infinite versus finite configurations, and \textit{(ii)} the dispersion analysis is based on the assumption of linear-elastic material behavior.

For a proper numerical evaluation of wave transmission, we calculated the forced vibration responses of a finite-sized structure with a design based on \text{Kelvin} unit cells twisted by 45$^\circ$. The CAD-based models of the fabricated structures were discretized using approximately $1.5\cdot 10^5$~quadratic tetrahedral elements, providing more than 10 elements per minimum shear wavelength at the highest analyzed frequency (8\,kHz). A frequency-domain analysis was conducted in \text{COMSOL Multiphysics}. The harmonic displacement  $u_{\mathrm{in}} = u_{0,z} e^{i \omega t}$ was prescribed in the center of the bottom plate, while all other boundaries were traction-free. The inertia of the screw junction used in the tests was accounted for by adding a concentrated mass on the excitation side. The axial displacements were extracted at the center of the top plate on the opposite side of the structure, and the transmission was computed as
\begin{equation}
T(\omega) = 20 \log_{10} \left( \frac{\overline{u}_{z,\mathrm{out}}}{\overline{u}_{z,\mathrm{in}}} \right),
\end{equation}
where $\overline{u}_{z}$ denotes the displacement averaged on the surface of a respective plate.

\begin{figure}[]
     \centering
     \begin{subfigure}[t]{0.49\textwidth}
         \centering
         \includegraphics[width=.95\textwidth]{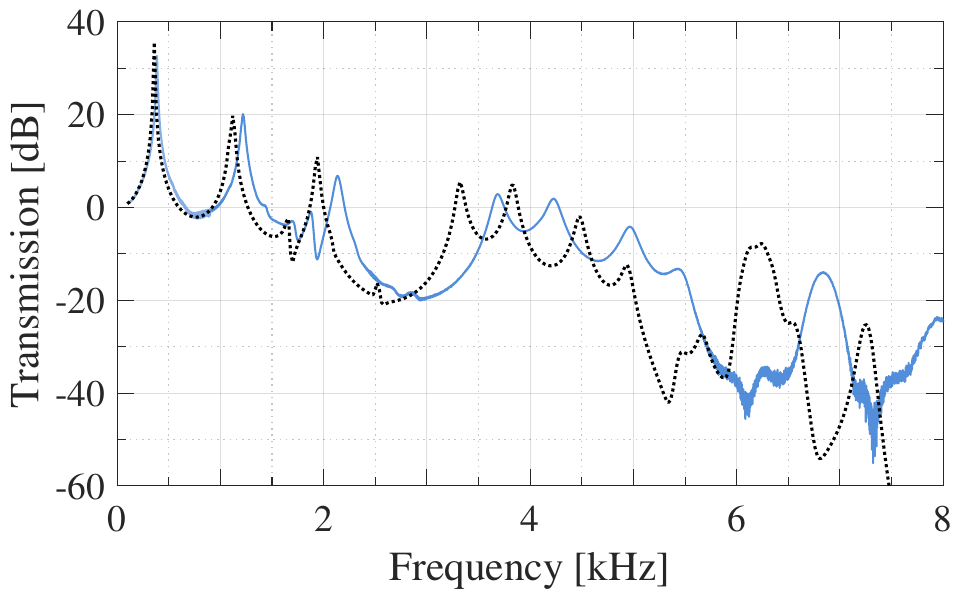}
         \caption{}\label{fig:LinElast}
     \end{subfigure}
     \hfill
     \begin{subfigure}[t]{0.49\textwidth}
         \centering
         \includegraphics[width=.95\textwidth]{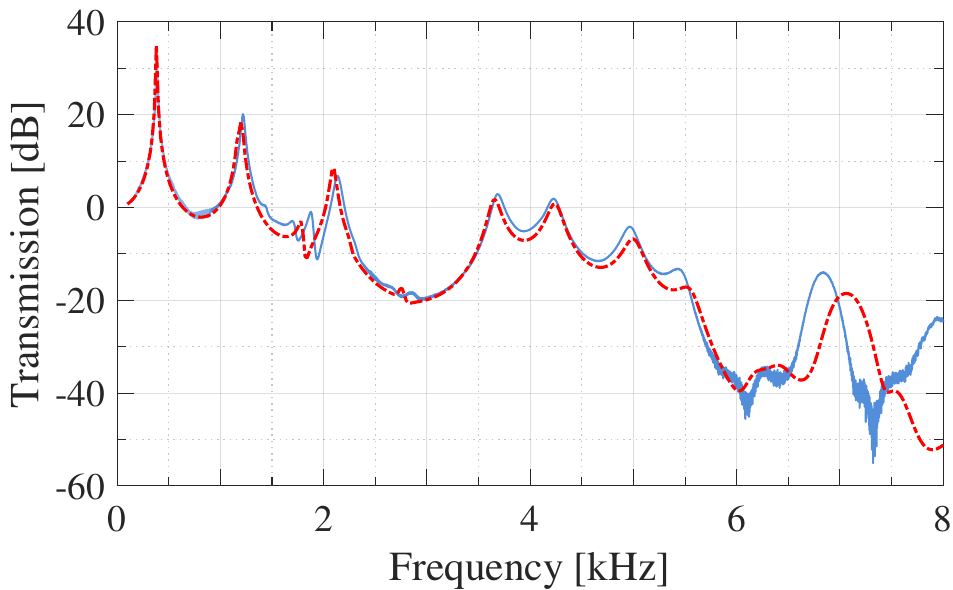}
         \caption{}\label{fig:FreqDepend}
     \end{subfigure}
    \caption{The measured (solid blue) and simulated (dotted and dashed) transmission spectra for a three-cell Kelvin chain twisted by $\theta_{z{+}}^{\hat{z}} = 45^\circ$. The simulations are performed assuming a (a) linear-elastic and (b) frequency-dependent viscoelastic material model. In the viscoelastic model, the Young’s modulus and the loss factor vary linearly with frequency according to Eq.~\eqref{eq:materialCorrection}.}\label{fig:Simulation}
\end{figure}

Figure~\ref{fig:LinElast} presents the numerical transmission spectrum computed under the assumption of linear-elastic material behavior, along with the experimental results, showing reasonable agreement at low frequencies below 1~kHz. However, the first attenuation band centered around 2~kHz is underestimated; at higher frequencies, the discrepancy increases further, particularly near the second and third transmission dips, resulting in significant deviations from measured values.

The observed mismatch can be attributed to two primary sources:
\begin{enumerate}
    \item Excitation imperfections: The numerical model assumes an ideal, axisymmetric excitation, while the experimental excitation inevitably involves off-axis components due to non-ideal positioning, imperfect gluing, or manufacturing inhomogeneities leading to the activation of additional, possibly localized, modes, particularly at higher frequencies.
    \item Viscous material losses: The base material -- resin -- has inherent dissipation causing frequency-dependent mechanical behavior, not captured by the linear-elastic model. Previous studies~\cite{Krushynska.2021b, Arretche.2019, DAlessandro.2018, Lewinska.2017} have shown that neglecting even small viscous losses leads to inaccurate transmission predictions at higher frequencies.
\end{enumerate}

To address the second issue, we implemented a simple empirical viscoelastic model with a single parameter to account for the frequency dependence in the \text{Young}'s modulus and the structural loss factor, following Ref.~\citep{Arretche.2019}:
\begin{align}
E(f) = E_0 + 100f \, [\mathrm{kPa/Hz}], \qquad \eta(f) = \eta_0 + 1\cdot10^{-6}f \, [1/\mathrm{Hz}],
\label{eq:materialCorrection}
\end{align}
where $E_0$ and $\eta_0$ correspond to the nominal linear-elastic parameters. The shown parameter values were adjusted to achieve closer agreement between the numerical and experimental transmission data. 
Keeping the same boundary and excitation conditions, the nominal material density was reduced by 10\% to account for discrepancies between the manufacturer's specifications and the actual behavior of the 3D-printed material. These include imperfections in the effective volume (e.g., menisci at strut connections) and slight density variations introduced by the 3D printing process.
\begin{figure}[]
     \centering
     \begin{subfigure}[t]{0.49\textwidth}
         \centering
         \includegraphics[width=.99\textwidth]{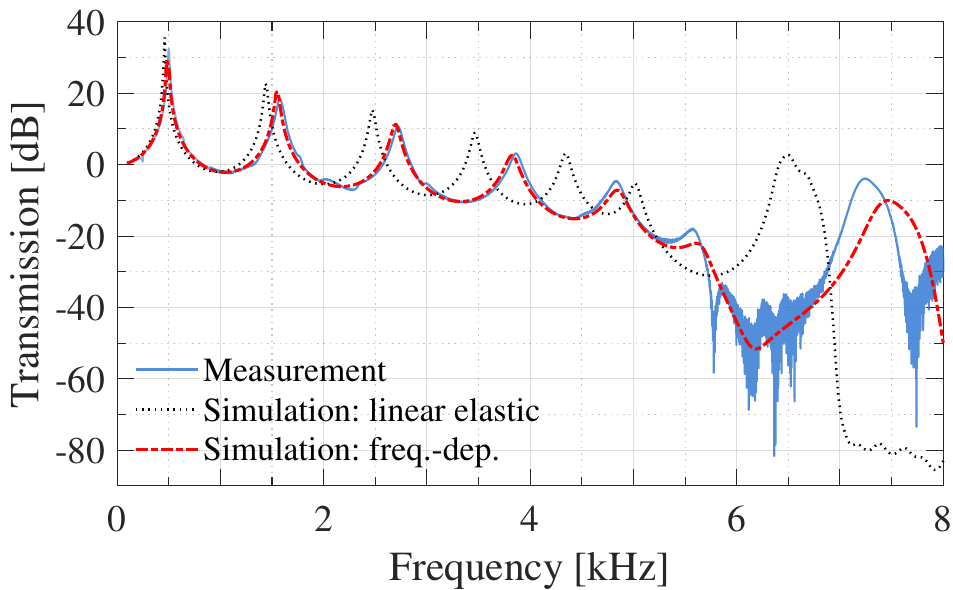}
         \caption{}\label{fig:TransSim_KCRef}
     \end{subfigure}
     \hfill
     \begin{subfigure}[t]{0.49\textwidth}
         \centering
         \includegraphics[width=.99\textwidth]{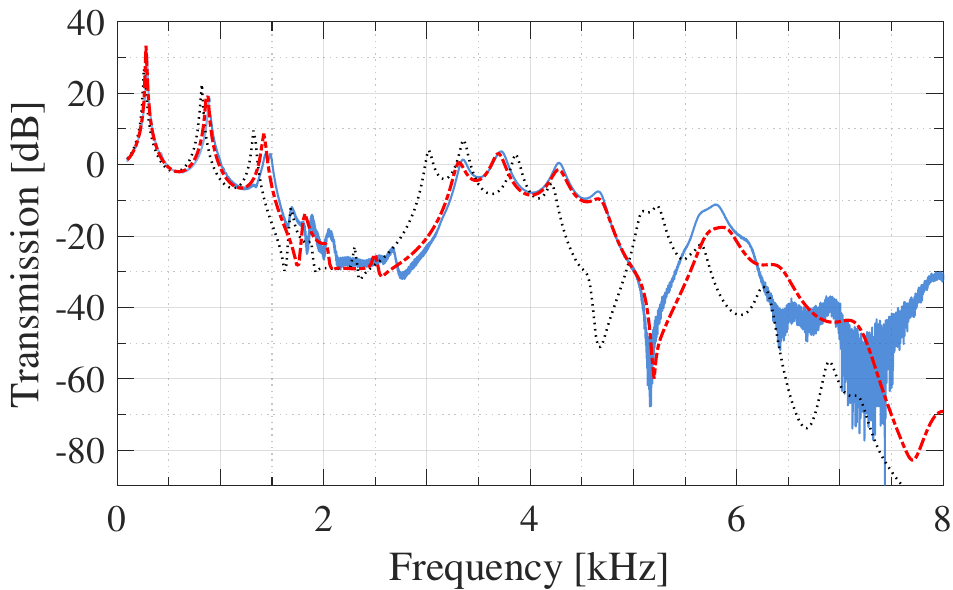}
         \caption{}\label{fig:TransSim_Twist90}
     \end{subfigure}
    \caption{The measured (solid blue) and simulated (dotted and dashed) transmission spectra for the three-cell reference and $\theta_{z{+}}^{\hat{z}} = 90^\circ$ twisted Kelvin chains. The black-dotted and red-dashed curves correspond to a linear-elastic and visco-elastic (Eq.~\eqref{eq:materialCorrection}) material model, respectively. The parameters were adjusted using the specimen shown in Fig.~\ref{fig:Simulation} and subsequently applied unchanged to all chain configurations, demonstrating the material model's generalizability.}
\label{fig:Trans_KCRef}
\end{figure}

Figure~\ref{fig:FreqDepend} presents the numerical transmission based on the frequency-dependent material behavior compared to the measured spectrum. The results show excellent agreement in the analyzed frequency range, with the simulation accurately capturing the first and second transmission dips near 2\,kHz and 5.5~kHz, respectively. However, the simulations overestimate the opening of the third dip at approximately 7~kHz, indicating that the assumed linear dependence of the moduli on frequency is insufficient to correctly reproduce the structural response at higher frequencies. 

The validity of the proposed frequency-dependent model was further validated in simulations for finite-size structures formed by reference \text{Kelvin} cells and those twisted by $90^\circ$.  Crucially, the material parameters calibrated for the $45^\circ$-case were applied to these distinct topologies without any further modification. Figure~\ref{fig:Trans_KCRef} shows that the frequency-dependent model accurately predicts the measured transmission spectra in both cases, showing particularly good agreement in the attenuation regions, without requiring any further adjustment of the parameters, demonstrating the high degree of generalizability of our viscoelastic framework across different lattice architectures. In contrast, the linear-elastic model fails to reproduce the measured spectra at higher frequencies. These results validate the viscoelastic material model and demonstrate its applicability to different unit-cell geometries made of the same constituent material. However, note that at frequencies above $\SI{6}{\kilo\hertz}$ the model cannot accurately capture the transmission behavior, underlying the need for more advanced descriptions of the material response, in agreement with other studies~\cite{Rouleau.2016, Cuenca.2014, DAlessandro.2018, Krushynska.2021}.

In summary, our findings reveal that the deviations of the linear-elastic simulations from the experimental results primarily arise from the frequency-dependent characteristics of the constituent resin, proving that incorporating material-intrinsic dissipation is critical for the accurate prediction of band-gap frequency bounds in polymer-based lattices. Possible manufacturing defects and excitation imperfections have a negligible impact and can be safely ignored without compromising accuracy.

%% file: content/05_Conclusion.tex
\section{Conclusion}
This work introduces a minimalist, topology- and mass-preserving approach to engineer wave-propagation characteristics in cellular metamaterial chains by introducing a single-parameter twist to the canonical Kelvin unit cell. By breaking the octahedral symmetry of this template geometry, we achieved chiral, anisotropic architectures that activate and tune band-gap frequencies without the need for the excessive geometrical complexity or added structural detail that often characterize current trends in the field.

The dispersion analysis revealed that this twist-induced symmetry breaking activates two distinct attenuation mechanisms: a wide \text{Bragg}-type band gap due to the structural periodicity and a narrower symmetry-induced gap arising from the coupling between longitudinal and torsional modes. Critically, both mechanisms are achieved by twisting only a single face of the Kelvin cell, and do not require embedded resonators or added bulk mass, thereby offering a straightforward and scalable approach to vibration control in lightweight structures. 

The numerical dispersion study was complemented by a physics-driven, analytical mass-spring model that not only qualitatively reproduces the \text{Bragg}-type band gaps but also captures the mode-coupling-induced avoided crossings that define the dynamics of twisted \text{Kelvin} cells. 

The practical feasibility of the proposed approach was verified via transmission measurements on finite-size SLA-manufactured specimens. The twisted structures with only three unit cells exhibited clear attenuation bands, with transmission dips up to 20\,dB. 

Furthermore, we demonstrated that accounting for the linear viscoelasticity of the base polymer is a fundamental requirement, rather than a secondary detail, for achieving predictive fidelity. The inclusion of frequency-dependent material properties was vital to align numerical predictions with experimental results, particularly at higher frequencies where idealized linear-elastic models typically fail.

In summary, the proposed twisting approach is versatile and easily parameterized to engineer wave-filtering behavior in architected lattice materials. While this study addressed 1D chains, future work will focus on expanding the design space to multi-directional periodicity, where the role of anisotropy in wave control is especially relevant. 

The kinematics of the coupled longitudinal--torsional behavior present a promising opportunity to amplify inertial effects, potentially enabling the formation of broader or multi-modal band gaps without added mass. Further refinement of the analytical model to capture such multi-directional and multi-modal interactions could provide deeper insights into these phenomena. In addition, strategically interrupting selected lattice-strut connections may provide a means to engineer defect states and achieve targeted energy localization within the structure. 

These research avenues, combined with improved predictive fidelity through detailed material characterization and multiphysics modeling, support the advancement of lattice architectures for applications in lightweight vibration isolation, reconfigurable structural systems, and wave-based sensing technologies. Beyond its wave-control capabilities, the adaptability of the proposed approach offers a versatile testbed for reduced-order and homogenized descriptions, including dynamic homogenization methods, such as higher-order gradient formulations~\cite{Molavitabrizi.2023} and extended continuum theories~\cite{Chen.2020}, as well as parametric model order reduction schemes~\cite{ReschSchopper.2026}.

%% file: content/06_a_Appendix_Methods.tex
\section{Dispersion eigenvalue problems}\label{Appendix:Methods}
In Sec.~3 of the manuscript, the dispersion eigenvalue problem is formulated and solved using both the $\omega(k)$ and the $k(\omega)$ approaches. This section provides additional details on how each formulation is implemented in \text{COMSOL} Multiphysics\textsuperscript{\textregistered}. For a detailed discussion of the theoretical background underlying both approaches, [\textsc{Laude (2020)}] is recommended.

\subsection{Dispersion eigenvalue problem -- $\omega (k) $}
The $\omega(k)$ formulation, also known as the \textit{indirect} approach, is readily available in the Solid Mechanics module and imposes Bloch--Floquet-type boundary conditions,
\begin{align}
  \bm{u}_\mathrm{dst} =
  \bm{u}_\mathrm{src}
  \mathrm{e}^{-\mathrm{i}\,\bm{k}\cdot(\bm{r}_\mathrm{dst}-\bm{r}_\mathrm{src})},
\end{align}
on the respective source and destination faces of the unit cell. By construction, the wavenumber is real-valued ($k \in \mathbb{R}$), and the resulting eigenvalues~$\lambda = -\mathrm{i} \omega $ are complex-valued. The underlying elastodynamic problem is solved subject to these Floquet-type boundary conditions while sweeping the wavevector $\bm{k} = [0, 0, k_z]^\mathrm{T}$ over the first Brillouin zone. The resulting eigenvectors represent modes propagating to infinity in space, whereas the imaginary part of the eigenvalue corresponds to attenuation (or amplification) of the vibrations in time.

\subsection{Dispersion eigenvalue problem -- $k (\omega) $}
As an alternative to the $\omega(k)$ approach, the dispersion eigenvalue problem can be formulated by prescribing real-valued frequencies, $\omega \in \mathbb{R}$, and solving for the admissible complex-valued wavenumbers, $k \in \mathbb{C}$, along a specified wavevector direction, here chosen as $\bm{k} = k_z \bm{e}_z$. In contrast to the former case, this formulation reveals all types of spatial wave solutions, including propagating, evanescent, and attenuated/amplified modes, as well as their mutual conversion along the course of dispersion. This enables an in-depth investigation of band-gap mechanisms and modal coupling phenomena.

To obtain the $k(\omega)$ formulation, the underlying elastodynamic problem is first considered and Bloch-type solutions are imposed to derive the strong form of the Bloch eigenvalue problem. This problem then serves as a template that is subsequently mapped onto the Coefficient Form PDE interface available in \text{COMSOL} Multiphysics\textsuperscript{\textregistered}. The following procedure outlines how the governing equations, boundary conditions, and Bloch-periodic constraints are formulated and implemented within the Coefficient Form PDE interface to compute the complex-valued dispersion relations, building on the initial contributions of [\textsc{Wang, Laude}~\textit{et al.}~\textsc{(2015)}] and [\textsc{Collet, Ouisse}~\textit{et al.}~\textsc{(2011)}] . Explicit expressions for the coefficient matrices are provided in Sec.~\ref{App:CoefficientMatrices}. 

\subsubsection{Governing Wave Equation and \text{Bloch} Solutions}
Introducing the \text{Lamé} parameters to describe the linear-elastic material behavior,
\begin{align}
    \mu = \frac{E}{2(1+\nu)}, 
    \qquad 
    \lambda = \frac{E\nu}{(1+\nu)(1-2\nu)},
    \label{eq:LameParams}
\end{align}
the time-harmonic \text{Navier--Cauchy} wave equation reads
\begin{align}
    (\lambda + \mu)\,\nabla(\nabla\!\cdot\!\bm{u})
    + \mu\,\nabla^2\bm{u}
    + \rho\,\omega^2\,\bm{u}
    = \bm{0}.
    \label{eq:WaveEq_Lame}
\end{align}
Inside the unit cell domain~$\Omega_\mathrm{C}$, solutions to Eq.~\eqref{eq:WaveEq_Lame}
take the \text{Bloch} form
\begin{align}
    \bm{u}(\bm{r},\bm{k})
    = \tilde{\bm{u}}_{k}(\bm{r})\,\mathrm{e}^{-\mathrm{i}\bm{k}\cdot\bm{r}}.
\end{align}
The case of one-dimensional periodicity is specified along the $\bm{e}_z$--direction by taking
\begin{align}
    \bm{k} = k_z \bm{e}_z, 
    \qquad 
    \bm{r} = z \bm{e}_z, 
    \qquad 
    \bm{u}(\bm{r}) = \tilde{\bm{u}}_k\,\mathrm{e}^{-\mathrm{i}k_z z},
\end{align}
Accordingly, the gradient of the displacement field $\bm{u} = \left[u_x, u_y, u_z \right]^\mathrm{T}$ is 
\begin{align}
    \nabla \bm{u}   = \left( \nabla \tilde{\bm{u}}_k \right)\mathrm{e}^{-\mathrm{i} k_z z} + \tilde{\bm{u}}_k \otimes\nabla \mathrm{e}^{-\mathrm{i} k_z z} = \mathrm{e}^{-\mathrm{i} k_z z} \left[ \nabla \tilde{\bm{u}}_k - \mathrm{i} k_z \tilde{\bm{u}}_k \otimes \bm{e}_z \right].
\end{align}
From this expression, the derivatives of the \text{Bloch} solution can be expressed compactly as
\begin{align}
    \nabla
    \;\rightarrow\;
    \nabla - \mathrm{i}k_z\bm{e}_z
    \label{eq:BlochOperator}
\end{align}
which permits the subsequent formulations:
\begin{equation}
\label{eq:BlochOperators}
\begin{aligned}
\text{Divergence} \quad
&\nabla\!\cdot\!\bm{u}
&\rightarrow\;&
\big(\nabla - \mathrm{i}k_z \bm{e}_z \big)\!\cdot\!\tilde{\bm{u}}_k
  = \nabla\!\cdot\!\tilde{\bm{u}}_k
  - \mathrm{i}k_z\,\tilde{u}_{k,z},
\\[6pt]
\text{Gradient of divergence}\quad
&\nabla(\nabla\!\cdot\!\bm{u})
&\rightarrow\;&
\big(\nabla - \mathrm{i}k_z \bm{e}_z\big)
  \!\big(\nabla\!\cdot\!\tilde{\bm{u}}_k - \mathrm{i}k_z\,\tilde{u}_{k,z}\big)
\\
&&&=
  \nabla(\nabla\!\cdot\!\tilde{\bm{u}}_k)
  - \mathrm{i}k_z\,\nabla\tilde{u}_{k,z}
  - \mathrm{i}k_z\,\bm{e}_z\,(\nabla\!\cdot\!\tilde{\bm{u}}_k)
  - k_z^2\,\tilde{u}_{k,z}\,\bm{e}_z,
\\[6pt]
\text{Laplacian}\quad
&\nabla^2\bm{u}
&\rightarrow\;&
\nabla^2\tilde{\bm{u}}_k
  - 2\,\mathrm{i}k_z\,\partial_z\tilde{\bm{u}}_k
  - k_z^2\,\tilde{\bm{u}}_k.
\end{aligned}
\end{equation}

Applying the \text{Bloch} operators~\eqref{eq:BlochOperators}
to the governing wave equation~\eqref{eq:WaveEq_Lame}
yields the strong form of the \text{Bloch} wave equation
for one-dimensional periodicity along $z$ in terms of~$\tilde{\bm{u}}_k$,
\begin{equation}
\begin{aligned}
&(\lambda+\mu)\Big[
    \nabla\!\big(\nabla\!\cdot\!\tilde{\bm{u}}_k\big)
    - \mathrm{i}k_z\,\nabla \tilde{u}_{k,z}
    - \mathrm{i}k_z\,\bm{e}_z\,\big(\nabla\!\cdot\!\tilde{\bm{u}}_k\big)
    - k_z^{2}\,\bm{e}_z\,\tilde{u}_{k,z}
\Big] \\
&\qquad
+\,\mu\Big[
    \nabla^{2}\tilde{\bm{u}}_k
    - 2\,\mathrm{i}k_z\,\partial_{z}\tilde{\bm{u}}_k
    - k_z^{2}\tilde{\bm{u}}_k
\Big]
+ \rho\,\omega^{2}\,\tilde{\bm{u}}_k
= \bm{0}.\label{eq:ExtendedBlochWaveEq}
\end{aligned}
\end{equation}

Through algebraic manipulation, this expression can be grouped according to the powers of $k_z$ as
\begin{align}
&\Big[(\lambda+\mu)\nabla(\nabla\!\cdot\!\tilde{\bm{u}}_k)
+\mu\nabla^{2}\tilde{\bm{u}}_k\Big]
-\mathrm{i}k_z\Big[(\lambda+\mu)
\big(\nabla \tilde{u}_{k,z}+\bm{e}_z(\nabla\!\cdot\!\tilde{\bm{u}}_k)\big)
+2\mu\,\partial_{z}\tilde{\bm{u}}_k\Big] \notag\\[4pt]
&\quad
- k_z^{2}\Big[\mu\,\tilde{\bm{u}}_k
+(\lambda+\mu)\,\bm{e}_z\,\tilde{u}_{k,z}\Big]
+\rho\,\omega^{2}\,\tilde{\bm{u}}_k
= \bm{0}, 
\label{eq:BlochWaveEq_z}
\end{align}
which serves as a base equation to be brought into the Coeffcient Form PDE.

\subsubsection{Coefficient Form PDE}
The standard \text{COMSOL} template equation reads
\begin{align}
     \bm{e_A} \Lambda^2 \bm{U} - \bm{d_a} \Lambda \bm{U} + \nabla\cdot (-\bm{c} \nabla\bm{U} - \bm{\alpha} \bm{U} + \gamma) +\bm{\beta} \cdot \nabla\bm{U} + \bm{a} \bm{U} = 0 \; \mathrm{in}\; \Omega.
     \label{eq:Comsol_pde}
 \end{align}

To map the strong form equation onto the Coefficient Form PDE, Eq.~\eqref{eq:BlochWaveEq_z} is rewritten by introducing the eigenvalue~$\Lambda$ and the dependent field variables~$\mathbf{U}$ as 
\begin{align}
  \Lambda \coloneqq - \mathrm{i}k_z, \quad \tilde{\bm{u}}_k  \coloneqq \bm{U} = \left[U_x, U_y, U_z \right]^\mathrm{T}
\end{align}
such that
\begin{equation}
\label{eq:BlochWaveGrouped}
\begin{aligned}
&\underbrace{\Big[(\lambda+\mu)\nabla(\nabla\!\cdot\!\bm{U})
+\mu\nabla^{2}\bm{U}\Big]}_{\text{2nd-order derivatives}}
\;+\;
\underbrace{\textcolor{blue}{\Lambda}\Big[(\lambda+\mu)
\big(\nabla \bm{U}_z +\bm{e}_z(\nabla\!\cdot\!\bm{U})\big)
+2\mu\,\partial_{z}\bm{U}\Big]}_{\text{1st-order derivatives}} \\[4pt]
&\;+\;
\underbrace{-\textcolor{blue}{\Lambda^2}\Big[\mu\,\bm{U}
+(\lambda+\mu)\,\bm{e}_z\,U_z\Big]}_{\text{0th-order}}
\;+\;
\underbrace{\rho\,\omega^{2}}_\text{inertia}\,\bm{U}
=\bm{0}.
\end{aligned}
\end{equation}

Conformity with the template PDE~\eqref{eq:Comsol_pde} is subsequently obtained by factoring out $\bm{U}$ and conducting a coefficient comparison. The inertial contribution defines the matrix $\bm{a}$, 
\begin{align}
  \rho\,\omega^{2}\,\bm{U} \Leftrightarrow \underbrace{\rho\,\omega^{2} \mathbf{I}}_{\coloneqq \bm{a}} \,\bm{U}, 
\end{align}
while the zeroth-order derivative terms (i.e., $\Lambda^2$) supply the quadratic eigenvalue term:
\begin{align}
  \mu\,\bm{U} +(\lambda+\mu)\,\bm{e}_z\,U_z \Leftrightarrow \underbrace{\left(\mu \bm{I} + (\lambda + \mu)\, \bm{e}_z \otimes \bm{e}_z \right)}_{\coloneqq \bm{e}_A} \bm{U}.
\end{align}

Likewise, by considering $\nabla(\nabla \cdot \bm{U}) = \nabla\cdot \left( \nabla \bm{U}^\mathrm{T} \right)$ and $\nabla^2 \bm{U} = \nabla \cdot \left( \nabla \bm{U} \right) $, the second-order derivative term can be rearranged as 
\begin{align}
(\lambda+\mu)\nabla(\nabla\!\cdot\!\mathbf{U})
+\mu\nabla^{2}\mathbf{U}
&\;\Leftrightarrow\;
\nabla\!\cdot\! \underbrace{
\Big[\,(\lambda+\mu)\,\left( \nabla \bm{U}^\mathrm{T}\right) + \mu\,\left( \nabla \bm{U}\right) \Big]}_{\coloneqq  \mathbf{c}\nabla \bm{U}}
\end{align}
where the components of the fourth-order tensor~$\bm{c}$ are given by 
\begin{align}
  c_{ijkl} = (\lambda + \mu)\delta_{il}\delta_{jk} + \mu \delta_{ik}\delta_{jl}.
\end{align}

Lastly, the first-order derivatives of Eq.~\eqref{eq:BlochWaveGrouped} are to be mapped onto the template PDE~\eqref{eq:Comsol_pde}. Assuming spatially constant coefficients, the template equation may be rewritten as 
\begin{align}
\nabla \cdot (-\boldsymbol{\alpha}\boldsymbol{U})
+ \boldsymbol{\beta}\cdot \nabla \boldsymbol{U}
=
\sum_j (\boldsymbol{\beta}^{(j)}-\boldsymbol{\alpha}^{(j)})\partial_j\boldsymbol{U}
\label{eq:firstOrderTemplate2Map}
\end{align}
where
\(
\boldsymbol{\alpha}=\{\boldsymbol{\alpha}^{(j)}\}_{j\in\{x,y,z\}},
\;
\boldsymbol{\beta}=\{\boldsymbol{\beta}^{(j)}\}_{j\in\{x,y,z\}},
\)
and
\(
\boldsymbol{\alpha}^{(j)},\boldsymbol{\beta}^{(j)}\in\mathbb{R}^{3\times3}.
\)

It is convenient to write the first-order order contributions
\begin{align}
  \textcolor{blue}{\Lambda}\Big[(\lambda+\mu)
    \big(\nabla \bm{U}_z +\bm{e}_z(\nabla\!\cdot\!\bm{U})\big)
+2\mu\,\partial_{z}\bm{U}\Big]
\label{eq:FirstOrderDerivatives}
\end{align}
as a sum of matrices multiplying the direction derivatives $\partial_j \bm{U}, j \in \left\{ x, y, z \right\} $. Using
\begin{align}
  \nabla \bm{U}_j = \begin{bmatrix}
    \partial_x U_j \\
    \partial_y U_j \\
    \partial_z U_j \\
  \end{bmatrix}, 
  \;\;
    \partial_{j}\bm{U}
  = \begin{bmatrix}
    \partial_j U_x \\
    \partial_j U_y \\
    \partial_j U_z \\
  \end{bmatrix}, 
  \;\;
  \nabla \cdot \bm{U} = \sum_j \partial_j U_j, 
\end{align}
the terms inside the parenthesis of Eq.~\eqref{eq:FirstOrderDerivatives} may be expressed as
\begin{subequations}\label{eq:lambdaLinearContributions}
\begin{align}
\bullet\;&
(\lambda+\mu)\nabla U_z
=
(\lambda+\mu)
\begin{bmatrix}
\partial_x U_z\\
\partial_y U_z\\
\partial_z U_z
\end{bmatrix}
\nonumber\\
&=
(\lambda+\mu)\left(
\begin{bmatrix}
0 & 0 & 1\\
0 & 0 & 0\\
0 & 0 & 0
\end{bmatrix}
\partial_x \mathbf{U}
+
\begin{bmatrix}
0 & 0 & 0\\
0 & 0 & 1\\
0 & 0 & 0
\end{bmatrix}
\partial_y \mathbf{U}
+
\begin{bmatrix}
0 & 0 & 0\\
0 & 0 & 0\\
0 & 0 & 1
\end{bmatrix}
\partial_z \mathbf{U}
\right),
\label{eq:lambdaLinearContributions:a}
\\[1.0ex]
\bullet\;&
(\lambda+\mu)\,\bm{e}_z(\nabla\cdot \bm{U})
=
(\lambda+\mu)
\begin{bmatrix}
0\\
0\\
\sum_j \partial_j U_j
\end{bmatrix}
\nonumber\\
&=
(\lambda + \mu) \left(
\begin{bmatrix}
0 & 0 & 0\\
0 & 0 & 0\\
1 & 0 & 0
\end{bmatrix}
\partial_x \mathbf{U}
+
\begin{bmatrix}
0 & 0 & 0\\
0 & 0 & 0\\
0 & 1 & 0
\end{bmatrix}
\partial_y \mathbf{U}
+
\begin{bmatrix}
0 & 0 & 0\\
0 & 0 & 0\\
0 & 0 & 1
\end{bmatrix}
\partial_z \mathbf{U}
\right),
\label{eq:lambdaLinearContributions:b}
\\[1.0ex]
\bullet\;&
2\mu\,\partial_{z}\bm{U}
=
2 \mu 
\begin{bmatrix}
1 & 0 & 0\\
0 & 1 & 0\\
0 & 0 & 1
\end{bmatrix}
\partial_{z}\bm{U}.
\label{eq:lambdaLinearContributions:c}
\end{align}
\end{subequations}

Collecting Eqs.~\eqref{eq:lambdaLinearContributions:a}-\eqref{eq:lambdaLinearContributions:c} and grouping all terms by derivative direction gives
\begin{align}
  \mathcal{L}_1(\bm{U}) = \textcolor{blue}{\Lambda} \sum_{j \in \{x, y, z\}} \mathbf{B}^{(j)} \partial_j \bm{U}
\end{align}
with 
\begin{align}
  \mathbf{B}^{(x)} 
  = (\lambda + \mu)
  \begin{bmatrix}
  0 & 0 & 1\\
  0 & 0 & 0\\
  1 & 0 & 0
\end{bmatrix}, \quad
 \mathbf{B}^{(y)} 
  = (\lambda + \mu)
  \begin{bmatrix}
  0 & 0 & 0\\
  0 & 0 & 1\\
  0 & 1 & 0
\end{bmatrix}, \quad
 \mathbf{B}^{(z)} =
  \begin{bmatrix}
  2\mu & 0 & 0\\
  0 & 2\mu & 0\\
  0 & 0 & 2(\lambda + 2\mu)
\end{bmatrix}
\label{eq:B_matrices}, 
\end{align}

which allows for direct coefficient-wise comparison with the first-order terms in Eq.~\eqref{eq:FirstOrderDerivatives} by imposing
\begin{align}
  \boldsymbol{\beta}^{(j)}-\boldsymbol{\alpha}^{(j)} = \textcolor{blue}{\Lambda} \mathbf{B}^{(j)}, \qquad
  j \in \{ x,y,z\}.
  \label{eq:MatchFirstOrder}
\end{align}
The choice of $\boldsymbol{\alpha}^{(j)}$ and $\boldsymbol{\beta}^{(j)}$ is constrained by the definition of the boundary conditions and therefore not arbitrary.

\subsubsection{Boundary Conditions}
Periodicity is enforced by imposing continuity conditions on the corresponding source and destination faces of the unit cell. Specifically, the displacement field and the associated flux are required to satisfy
\begin{align}
\boldsymbol{U}_{\mathrm{dst}} &= \boldsymbol{U}_{\mathrm{src}},
\qquad \mathrm{and} \;\;
\boldsymbol{n}\cdot\bigl(\boldsymbol{J}_{\mathrm{dst}}-\boldsymbol{J}_{\mathrm{src}}\bigr) =0,
\end{align}
where~$\bm{n}$ denotes the outward normal vector of the source boundary and~$\boldsymbol{J}$ the flux term defined as
\begin{align}
\boldsymbol{J} \coloneqq -\boldsymbol{c}\nabla\boldsymbol{U}
-\boldsymbol{\alpha}\boldsymbol{U}.
\end{align}
Consequently, $\boldsymbol{\alpha}^{(j)}$ is fixed by its contribution to the Bloch-modified flux resulting from the shifted gradient operator introduced in Eq.~\eqref{eq:BlochOperator}. Once 
all~$\boldsymbol{\alpha}^{(j)}$ are determined in this manner, the matrices~$\boldsymbol{\beta}^{(j)}$ follow directly from coefficient matching of the remaining first-order terms in Eq.~\eqref{eq:MatchFirstOrder}.

\subsubsection{Final Coefficient Matrices for Numerical Implementation}\label{App:CoefficientMatrices}
The following expressions summarize the coefficient matrices used to compute the complex-valued dispersion curves shown in Sec.~3 of the manuscript. All results were obtained by solving the resulting quadratic eigenvalue problem using the \textsc{Arpack} eigensolver, with the frequency~$\omega$ prescribed as sweeping parameter and the Bloch parameter~$\Lambda = -\mathrm{i} k_z$ treated as the eigenvalue. The coefficient matrices listed below are sufficient to reproduce the reported dispersion relations and stated as implemented in COMSOL’s Coefficient-Form PDE. Apparent sign differences with respect to the strong-form derivation (including terms involving $\bm{a}, \bm{e_A}$ and $\bm{c}$) arise from the flux definition  and the time-harmonic formulation.
\begin{align}
  \bm{a} =
\begin{bmatrix}
-\rho\,\omega^{2} & 0 & 0\\[3pt]
0 & -\rho\,\omega^{2} & 0\\[3pt]
0 & 0 & -\rho\,\omega^{2}
\end{bmatrix}, \;\;
\bm{e}_A =
\begin{bmatrix}
-\mu & 0 & 0\\[3pt]
0 & -\mu & 0\\[3pt]
0 & 0 & -(\lambda+2\mu)
\end{bmatrix}
\end{align}

\begin{align}
\mathbf{c} = &
\begin{bmatrix}
  \mathbf{C}_{xx} & \mathbf{C}_{xy} & \mathbf{C}_{xz} \\
  \mathbf{C}_{yx} & \mathbf{C}_{yy} & \mathbf{C}_{yz} \\
  \mathbf{C}_{zx} & \mathbf{C}_{zy} & \mathbf{C}_{zz} \\
\end{bmatrix}
\qquad \text{with}
\end{align}

\begin{align}
\nonumber
\mathbf{C}_{xx} &= 
\begin{bmatrix}
\lambda+2\mu & 0 & 0 \\
0 & \mu & 0 \\
0 & 0 & \mu
\end{bmatrix},
&
\mathbf{C}_{xy} &=
\begin{bmatrix}
0 & \lambda & 0 \\
\mu & 0 & 0 \\
0 & 0 & 0
\end{bmatrix},
&
\mathbf{C}_{xz} &=
\begin{bmatrix}
0 & 0 & \lambda \\
0 & 0 & 0 \\
\mu & 0 & 0
\end{bmatrix},
\\ \nonumber
\mathbf{C}_{yx} &=
\begin{bmatrix}
0 & \mu & 0 \\
\lambda & 0 & 0 \\
0 & 0 & 0
\end{bmatrix},
&
\mathbf{C}_{yy} &=
\begin{bmatrix}
\mu & 0 & 0 \\
0 & \lambda+2\mu & 0 \\
0 & 0 & \mu
\end{bmatrix},
&
\mathbf{C}_{yz} &=
\begin{bmatrix}
0 & 0 & 0 \\
0 & 0 & \lambda \\
0 & \mu & 0
\end{bmatrix},
\\ \nonumber
\mathbf{C}_{zx}&=
\begin{bmatrix}
0 & 0 & \mu \\
0 & 0 & 0 \\
\lambda & 0 & 0
\end{bmatrix},
&
\mathbf{C}_{zy}&=
\begin{bmatrix}
0 & 0 & 0 \\
0 & 0 & \mu \\
0 & \lambda & 0
\end{bmatrix},
&
\mathbf{C}_{zz} &=
\begin{bmatrix}
\mu & 0 & 0 \\
0 & \mu & 0 \\
0 & 0 & \lambda+2\mu
\end{bmatrix}.
\end{align}

\begin{align}
\mathbf{\alpha} = \textcolor{blue}{-\Lambda} 
\begin{bmatrix}
(0,0,\mu ) &
(0,0,0) &
\big(\lambda,\,0,\,0\big)
\\[6pt]
(0,0,0) &
(0,0,  \mu) &
\big(0,\,\lambda,\,0\big)
\\[6pt]
\big( \mu,\,0,\,0 \big) &
\big(0,\, \mu,\,0\big) &
\big(0,\,0,\, \lambda+2\mu\big)
\end{bmatrix}, \;\;
\mathbf{\beta} = \textcolor{blue}{\Lambda} 
\begin{bmatrix}
(0,\;0,\;\mu)
&
(0,\;0,\;0)
&
(\mu,\;0,\;0)
\\[6pt]
(0,\;0,\;0)
&
(0,\;0,\;\mu)
&
(0,\;\mu,\;0)
\\[6pt]
(\lambda,\;0,\;0)
&
(0,\;\lambda,\;0)
&
(0,\;0,\; \lambda+ 2\mu)
\end{bmatrix}
\end{align}

%% file: content/06_b_Appendix_AnalyticalModel.tex
\section{ Matricees \& Parameters}\label{Appendix:Analytical}
\subsection{Combined longitudinal–torsional and flexural model: Mono-atomic case}
The dispersion relation shown in Fig.~4a of the manuscript is obtained by enforcing the vanishing determinant condition on the dynamic stiffness matrices defined in Eq.~(12) and~Eq.~(16) of the manuscript. Following the estimation of the effective phase velocities based on the edge modes of the acoustic branches, the effective stiffness is evaluated as
\begin{align}
    k_\mathrm{eff} = m_\mathrm{eff} \left( \frac{c_\mathrm{ph}}{a} \right)^2,
\end{align}
where $m_\mathrm{eff}$ denotes the static mass of the unit cell, and $a$ represents the characteristic lattice spacing.
In summary, the following parameters were implemented in the dynamic stiffness matrix to obtain the dispersion relation.

\begin{table}[h!]
\centering
\caption{Model parameters for the mono-atomic combined longitudinal–torsional and flexural model. The corresponding dispersion relation is shown in Fig.~4a in the manuscript.}\label{tab:model_parameters}
\begin{tabular}{llll}
\toprule
Parameter & Value & Unit & Description \\
\midrule
$a$ & $0.02$ & $\mathrm{m}$ & Unit-cell length \\
$m_\mathrm{eff}$ & $0.471\cdot10^{-3}$ & $\mathrm{kg}$ & Static mass of a single Kelvin unit cell (from COMSOL) \\
$J$ & $m_\mathrm{eff}$ & $\mathrm{kg\,m^2}$ & Rotational inertia \\
$I_{yy}$ & $m_\mathrm{eff}/8$ & $\mathrm{m^4}$ & Area moment of inertia \\
$K_\mathrm{l}$ & $132\cdot10^{3}$ & $\mathrm{N/m}$ & Longitudinal stiffness \\
$K_\mathrm{t}$ & $30\cdot10^{3}$ & $\mathrm{N\,m/rad}$ & Torsional stiffness \\
$K_\mathrm{s}$ & $25\cdot10^{3}$ & $\mathrm{N/m}$ & Shear stiffness (for bending) \\
$K_\mathrm{b}$ & $8\cdot10^{3}$ & $\mathrm{N\,m/rad}$ & Rotational bending stiffness\\
$K_\mathrm{lt}$ & $0$ & $-$ & Compression–torsion coupling coefficient \\
$K_\mathrm{sb}$ & $0.55$ & $-$ & Shear–rotation coupling coefficient \\
\bottomrule
\end{tabular}
\end{table}

\subsection{Longitudinal–torsional model: Di-atomic case}
To obtain the dispersion relation shown in Fig.~6a of the manuscript, the mass–spring model is extended to a diatomic configuration (cf. manuscript Fig.~3a), thus featuring four degrees-of-freedom~$\mathbf{U} = \left[u_1, \varphi_1, u_2, \varphi_2 \right]^\mathrm{T}$. The dispersion relation is obtained by solving
\begin{align}
    \det\left[ \mathbf{K}(q) -\omega^2 \mathbf{M}\right] = 0,
\end{align}
where $\mathbf{M}=\operatorname{diag}\!\big(m_1,\;J_1,\;m_2,\;J_2\big)$ and $\mathbf{K}(q)$ the assembled stiffness matrix that is composed diverging axial and rotational stiffnesses~$K_{\mathrm{l}_{1,2}}$ and $K_{\mathrm{t}_{1,2}}$ and the elastic longitudinal--torsional coupling stiffness~$K_{\mathrm{lt}}$ such that
\begin{align}
    \mathbf{K}(q) = \mathbf{K}_0(q) + K_\mathrm{lt} \mathbf{K}_c(q)
\end{align}
with

\begin{equation}
\mathbf{K}_0(q) =
\begin{bmatrix}
K_{\mathrm{l},1}+K_{\mathrm{l},2} & 0 & -K_{\mathrm{l},1}-K_{\mathrm{l},2}e^{-iqa} & 0 \\[4pt]
0 & K_\mathrm{t,1}+K_\mathrm{t,2} & 0 & -K_\mathrm{t,1}-K_\mathrm{t,2}e^{-iqa} \\[4pt]
-K_{\mathrm{l},1}-K_{\mathrm{l},2}e^{iqa} & 0 & K_{\mathrm{l},1}+K_{\mathrm{l},2} & 0 \\[4pt]
 0 & -K_\mathrm{t,1}-K_\mathrm{t,2}e^{iqa} & 0 & K_\mathrm{t,1}+K_\mathrm{t,2}
\end{bmatrix},
\end{equation}

and
\begin{equation}
\mathbf{K}_c(q) =
\begin{bmatrix}
0 & 2 & 0 & -(1+e^{-iqa}) \\[4pt]
2 & 0 & -(1+e^{iqa}) & 0 \\[4pt]
0 & -(1+e^{iqa}) & 0 & 2 \\[4pt]
-(1+e^{-iqa}) & 0 & 2 & 0
\end{bmatrix}.
\end{equation}

To qualitatively reproduce the band-gap and avoided-crossing behavior, the following dimensionless parameters were implemented in the dynamic stiffness matrix.
\begin{table}[h!]
\centering
\caption{Parameters for the diatomic lumped mass–spring–torsion system. The corresponding band structure is shown in Fig.~6a of the manuscript}
\begin{tabular}{llll}
\toprule
Parameter & Value & Unit & Description \\
\midrule
$a$ 
& $0.02$ 
& $\mathrm{m}$ 
& Unit-cell length \\

$m_1$ 
& $4.71\cdot10^{-4}$ 
& $\mathrm{kg}$ 
& Static mass of a $45^\circ$-twisted Kelvin unit cell (from COMSOL) \\

$m_2$ 
& $0.75\,m_1$ 
& $\mathrm{kg}$ 
& Effective mass (second subcell) \\

$J_1$ 
& $0.5\,m_1$ 
& $\mathrm{kg\,m^2}$ 
& Effective rotational inertia \\

$J_2$ 
& $0.7\,J_1$ 
& $\mathrm{kg\,m^2}$ 
& Effective rotational inertia (second subcell) \\

$K_{\mathrm{l},1}$ 
& $8.5\cdot10^{4}$ 
& $\mathrm{N/m}$ 
& Longitudinal stiffness \\

$K_{\mathrm{l},2}$ 
& $0.6\,K_{\mathrm{l},1}$ 
& $\mathrm{N/m}$ 
& Longitudinal stiffness (second subcell) \\

$K_{\mathrm{t},1}$ 
& $3.0\cdot10^{4}$ 
& $\mathrm{N\,m/rad}$ 
& Torsional stiffness \\

$K_{\mathrm{t},2}$ 
& $0.7\,K_{\mathrm{t},1}$ 
& $\mathrm{N\,m/rad}$ 
& Torsional stiffness (second subcell) \\

$K_{\mathrm{lt}}$ 
& $5.0\cdot10^{3}$ 
& $-$ 
& Longitudinal–torsional coupling stiffness \\
\bottomrule

\end{tabular}\label{tab:diatomic_parameters}
\end{table}

%% file: content/06_c_Appendix_Experiments.tex
\section{Experimentally tested structures}\label{Appendix:Samples}

\begin{figure}[htbp]
    \centering
    \includegraphics[width = .7 \textwidth]{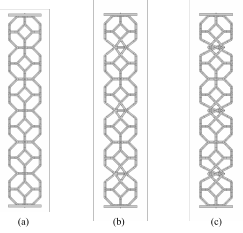}
    \caption{Overview of experimentally tested finite-periodic structures. (a) Reference Kelvin cell, 
    (b) twisted by~$\theta_j^{\hat{i}} = \pm 45^\circ$,
    (c) twisted by~$\theta_j^{\hat{i}} = \pm 90^\circ$.
    All lattices have identical dimensions. To highlight the geometrical differences, the structures are presented as 2D projections.
    }\label{fig:Supp_Specimens}
\end{figure}

In Sec.~4 of the manuscript, three finite periodic samples derived from the Kelvin cell are used to validate the proposed frequency-dependent material model. To highlight the geometrical differences, Fig.~\ref{fig:Supp_Specimens} presents 2D projections of the considered test cases. The lattice dimensions are identical for all structures.
